\documentclass[aps,floatfix,prb,reprint,notitle,
]{revtex4-2}

\usepackage{braket}
\usepackage{graphicx}
\usepackage{bm}

\usepackage{mathtools}

\usepackage{xcolor}
\usepackage{siunitx}
\usepackage{chemformula}
\usepackage{esint}
\usepackage{bbm}
\usepackage{amssymb}
\usepackage{tabularx}
\usepackage{xcolor}
\usepackage{tikz}

\definecolor{diagorange}{RGB}{240, 162, 15}
\definecolor{diagred}{RGB}{205, 0, 0}

\newcommand{\dline}{%
  \mathord{\tikz[baseline=-0.5ex, x=1em, y=1em]%
    \draw[thick] (0,0) -- (1.0,0);}}

\newcommand{\triangler}{%
  \mathord{\tikz[baseline=-0.5ex, x=1em, y=1em]%
    \fill[diagorange] (0,-0.40)--(0,0.40)--(0.80,0)--cycle;}}

\newcommand{\trianglel}{%
  \mathord{\tikz[baseline=-0.5ex, x=1em, y=1em]%
    \fill[diagorange] (0.80,-0.40)--(0.80,0.40)--(0,0)--cycle;}}

\newcommand{\dsquare}{%
  \mathord{\tikz[baseline=-0.5ex, x=1em, y=1em]%
    \fill[diagred] (0,-0.40) rectangle (0.80,0.40);}}

\newcommand{\doubleline}{%
  \mathord{\tikz[baseline=-0.5ex, x=1em, y=1em]{%
    \draw[thick] (0, 0.40) .. controls (0.33, 0.55) and (0.67, 0.55) .. (1.0, 0.40);
    \draw[thick] (0,-0.40) .. controls (0.33,-0.55) and (0.67,-0.55) .. (1.0,-0.40);}}}

\newcommand{\Xkvertex}{%
  \mathord{%
    \mkern-2mu%
    \tikz[baseline=-0.5ex, x=1em, y=1em, line cap=round, line join=round]{%
      \draw[thick] (-0.08,0) -- (0.25,0);
      \draw[thick] (0.60,0) circle[radius=0.35];
    }%
  }%
}

\newcommand{\Ykvertex}{%
  \mathord{%
    \tikz[baseline=-0.5ex, x=1em, y=1em, line cap=round, line join=round]{%
      \draw[thick] (0.70,0) circle[radius=0.35];
      \draw[thick]
        (-0.30,0.32) .. controls (0.05,0.32) and (0.22,0.18) .. (0.40,0.18);
      \draw[thick]
        (-0.30,-0.32) .. controls (0.05,-0.32) and (0.22,-0.18) .. (0.40,-0.18);
    }%
  }%
}

\usepackage{tikz}
\usetikzlibrary{decorations.pathmorphing,arrows.meta,shapes.geometric}
\usepackage{subfigure}
\newcommand{\eqname}{Eq.}
\newcommand{\secname}{Sec.}
\newcommand{\figname}{Fig.}

\def\XXint#1#2#3{{\setbox0=\hbox{$#1{#2#3}{\int}$}
     \vcenter{\hbox{$#2#3$}}\kern-.5\wd0}}

\DeclareMathOperator{\tr}{Tr}

\newcommand{\avg}[1]{\left\langle #1 \right\rangle_{\!\rho}}
\newcommand{\Dthree}{\stackrel{\text{\tiny(3)}}{\Phi}}
\newcommand{\Dfour}{\stackrel{\text{\tiny(4)}}{\Phi}}
\newcommand{\dthree}{\stackrel{\text{\tiny(3)}}{D}}
\newcommand{\dfour}{\stackrel{\text{\tiny(4)}}{D}}
\newcommand{\calL}{\mathcal{L}}
\newcommand{\calG}{\mathcal{G}}

\DeclareSIUnit\hartree{\protect \text {\protect \ensuremath {E}}_{\protect \mathrm {h}}}
\DeclareSIUnit\bohr{\protect \text {\protect \ensuremath {a}}_{0}}

\begin{document}

\preprint{APS/123-QED}

\title{Efficient simulation of second-order phase transitions\\ in quantum anharmonic materials}

\author{Andrea Baldanza and Lorenzo Monacelli}
 \affiliation{Dipartimento di Fisica, Universit\`a di Roma Sapienza}

\date{\today}

\begin{abstract}
    When a crystal undergoes a second-order structural phase transition, such as in ferroelectric, Peierls, and charge-density waves, the diverging fluctuations of the order parameter lead to the breakdown of the standard phonon quasiparticle picture. Simulating these highly anharmonic regimes is notoriously challenging, as methods such as molecular dynamics suffer from a critical slowdown near the transition point, while the harmonic approximation fails dramatically at saddle points of the energy landscape stabilized by quantum or thermal ionic fluctuations.
    
    This work introduces a new approach, based on the variational free-energy principle, to predict critical long-range behavior and dynamical spectra in strongly anharmonic systems, even when quantum ionic fluctuations dominate. The proposed framework builds upon the stochastic self-consistent harmonic approximation but reduces its computational scaling with the number of atoms, $N$, from $O(N^6)$ to $O(N^2)$ and the memory requirement from $O(N^4)$ to $O(N)$.
    We benchmark the method on the prototypical lead-free metal-halide perovskite \ch{CsSnI3}, a promising candidate for photovoltaic engineering, simulating its phase stability and Raman spectrum near the phase transition, where the breakdown of the quasiparticle picture becomes evident.  We demonstrate the effectiveness of the method by computing the full free-energy Hessian and the critical temperature in a supercell with $1080$ atoms. Such simulations would have required tens of thousands of years with the legacy approach; it is now feasible in a few hours on consumer hardware.
\end{abstract}

\maketitle

\section{Introduction}

Second-order structural phase transitions play a fundamental role in statistical physics and materials science, thanks to the peculiar properties of the critical transition point: a thermodynamic state where the excitation energy of the order parameter vanishes.
At this critical point, fluctuations in the order parameter grow with system size, and the system's response to perturbations involving the order parameter diverges.
This behavior is desirable for many technological applications as it results in a macroscopic response to small fields\cite{cochran_crystal_1960,cochran_dielectric_1962}. Moreover, the large fluctuations in the order parameter interact with the rest of the system, introducing strong phonon-phonon scattering. This is exploited in thermal transport, where critical fluctuations slow down heat propagation, resulting in a drop in thermal conductivity while retaining almost unchanged electronic properties. Prototypical examples include group IV-VI alloys such as \ch{SnTe}, \ch{GeTe}\cite{dangic_origin_2021,dangic_molecular_2022}, \ch{SnSe}\cite{aseginolaza_phonon_2019}, and \ch{SnS}\cite{aseginolaza_strong_2019}, some of which are among the most efficient bulk thermoelectric materials known today. In several of these materials, the criticality is hidden by quantum fluctuations of nuclei, as in the case of \ch{SrTiO3}\cite{verdi_quantum_2023,shin_quantum_2021}, \ch{KTaO3}\cite{ranalli_electron_2024}, and \ch{PbTe}\cite{romero_thermal_2015,ribeiro_strong_2018}, which serve as interesting playgrounds for nonlinear phononics\cite{xian_li_terahertz_2019,nova_metastable_2019,melnikov_anharmonic_2023,cheng_terahertz-driven_2023,libbi_nonequilibrium_2025}, where a ferroelectric state can be persistently induced on the picosecond timescale by applying a sufficiently strong mid-IR/THz pulse.

Simulating critical points is extremely challenging for two reasons: (i)~the correlation length of the order parameters decays polynomially, meaning that extremely large simulation cells are required to properly account for this divergence, and (ii)~the zero excitation energy of the order parameter corresponds to a divergent time period of its oscillation, which results in the well-known critical slowdown of molecular dynamics simulations\cite{souvatzis_entropy_2008}.
To overcome these issues, several mean-field approaches have been introduced\cite{hellman_lattice_2011,SCALID,QSCALID,knoop_tdep_2024,bottin_-tdep_2020}.
Among these, the self-consistent phonons\cite{tadano_self-consistent_2015} and the stochastic self-consistent harmonic approximation (SSCHA)\cite{errea_anharmonic_2014,monacelli_stochastic_2021} have encountered great success, as they also account for ionic quantum effects and do not rely on molecular dynamics (or path-integral molecular dynamics) trajectories affected by the aforementioned issues.
The key idea behind the SSCHA is to approximate the quantum ionic density matrix as the most general Gaussian (which includes interatomic covariances), and optimize its average position and covariance matrix to minimize the free energy.

In particular, the SSCHA can exploit crystal symmetries to compute the thermodynamic conditions at the critical point using the elegant Landau theory of phase transitions: we define a positional free energy $\mathcal F$ as a function of the order parameter $\Delta$, and assess the sign of the $\mathcal F$ second derivative around $\Delta=0$\cite{bianco_second-order_2017}.
This approach has been applied successfully to systems with strong quantum and thermal anharmonic fluctuations, including high-pressure hydrogen\cite{Borinaga2017,monacelli_black_2020,monacelli_quantum_2023}, hydrides and hydrates\cite{errea_quantum_2016,errea_quantum_2020,cherubini_quantum_2024,monacelli_hydrogen_2025}, thermoelectric materials\cite{aseginolaza_phonon_2019,aseginolaza_strong_2019,monacelli_first-principles_2023}, transition metal oxides\cite{verdi_quantum_2023,ranalli_temperature-dependent_2023} bending rigidity in graphene\cite{aseginolaza_bending_2024}, and it has unveiled the ionic entropic origin of charge-density wave melting in transition-metal dichalcogenides\cite{SkyZhou2020,bianco_quantum_2019,Diego2021}.

However, the calculation of the positional curvature of the free energy, i.e., the free energy Hessian, requires the knowledge of 3-phonon and 4-phonon scattering tensors, which are expensive to compute and memory-heavy to store without approximated compression techniques\cite{zhou_lattice_2014,togo_distributions_2015,zhou_compressive_2019,eriksson_hiphive_2019-1,wu_lattice_2023,lin_first-principles_2026}. As a result, the nonperturbative computational of the full free energy Hessian within the SSCHA scales as $O(N^6)$, with $N$ the number of atoms in the simulation cell, limiting its application to supercells with less than 50 atoms, too small to obtain converged results. Phase transitions have therefore been mostly investigated while ignoring 4-phonon scattering, using the so-called ``bubble'' approximation\cite{bianco_second-order_2017,monacelli_analyzing_2025}, which is considered state-of-the-art for nonperturbative phonon calculations.

In this work, we introduce two algorithmic innovations that drastically reduce the computational cost of inverting the high-rank anharmonic scattering tensors. The first is a reformulation of the free-energy Hessian problem as the inversion of a static Liouvillian operator via iterative solvers, completely bypassing the explicit construction of the fourth-order force-constant tensor. The second is a reformulation of the entire SSCHA linear response in momentum space. 
For the full Hessian, the resulting computational scaling is reduced from $\mathcal O(N^6) \to \mathcal{O}(N^2)$, while the memory requirement of the algorithm drops from $\mathcal O(N^4)\to \mathcal O(N)$, where $N$ indicates the number of atoms in the supercell.
Notably, for a $6\times6\times6$ supercell, we achieve a speedup on the order of $10^9$, enabling simulations that were simply impossible before. This dramatic speedup unlocks full mode mixing in free-energy Hessians and phonon spectral functions, bypassing the traditional bubble approximation at a fraction of the legacy computational cost.

In \secname~\ref{sec:hessian} we review the theoretical context of the free energy Hessian, its link to the Landau theory of phase transitions, and discuss the origin of the computational bottleneck in state-of-the-art implementations. Then, \secname~\ref{sec:linear:system} and \ref{sec:qspace} present the details of the new algorithm introduced in this work.
The method's scalability with system size in both time and memory is benchmarked on a simple gold crystal (FCC) and compared with the previous state of the art (\secname~\ref{sec:benchmark}). 
Then, we apply the method to a real-case material, \ch{CsSnI3}, computing the phase diagram and showing that it correctly identifies the phase transition. \\
The same formalism also applies to the evaluation of phonon spectral functions within the dynamical extension of the static SSCHA, the Time-Dependent Self-Consistent Harmonic Approximation (TD-SCHA)\cite{monacelli_time-dependent_2021, siciliano_wigner_2023}.
Therefore, we characterise the breakdown of the phonon quasiparticle picture in the simulated Raman spectrum of \ch{CsSnI3}, with a supercell exceeding 1000 atoms (\secname~\ref{sec:cssni3}).

\section{The free energy Hessian}
\label{sec:hessian}

The positional free-energy Hessian is the crucial quantity in the Landau theory of phase transitions for identifying a critical point\cite{bianco_second-order_2017}. 
The positional free energy $\mathcal F(\Delta)$ is a function of the order parameter whose curvature goes to zero at the critical temperature. A displacive phase transition is such that the order parameter can be expressed as a linear combination of the average atomic position:
$\Delta = \sum_a e_\Delta^a \left<r_a- r^{(0)}_a\right>$, where $\bm r$ is the vector of the Cartesian coordinates of each atom in the cell (the index $a$ run over both $x,y,z$ and atomic indices), $\bm {r^{(0)}}$ indicates the average positions of atoms in the high-symmetry (high temperature) phase, and $\bm {e_\Delta}$ is the atomic displacement pattern in the broken symmetry phase (bold symbols indicate vectors and matrices). Thus, $\Delta = 0$ for $T>T_c$, while the lattice centroids deform following $\bm {e_\Delta}$ below the phase transition.

The Landau free energy $\mathcal F(\Delta)$ can be rigorously defined in quantum mechanics\cite{monacelli_analyzing_2025} by constraining $\Delta$ on the equilibrium density matrix, while minimizing the global free energy functional under the Gibbs variational principle:
\begin{equation}
    \mathcal F(\Delta) = \min_{\hat \rho} \left\{F[\hat\rho] + \lambda\left(\avg{\sum_{a}e_\Delta^a(r_a - r^{(0)}_a)} - \Delta\right)\right\},
    \label{eq:F:delta:def}
\end{equation}
and $\lambda$ is chosen to impose the $\Delta$ constrain by making $\mathcal F(\Delta)$ stationary. Here, $\hat\rho$ is the trial density matrix and $F[\hat\rho]$ is the free energy functional defined
\begin{equation}
    F[\hat\rho] = \avg{\hat H} - TS[\hat\rho],
\end{equation}
$$
\avg{\hat H} = \tr\left[\hat\rho \hat H\right],\qquad
S[\hat \rho] = -k_b\avg{\ln\hat\rho},
$$
and $\hat H$ is the Born-Oppenheimer nuclear Hamiltonian.
In practice, if the specific symmetry-breaking pattern $\bm {e_\Delta}$ is unknown, we can define the positional free energy on all centroids $\tilde r_i$ as
\begin{equation}
    \mathcal F(\tilde r_1,\cdots ,\tilde r_n) = \min_{\hat \rho} \left[F[\hat\rho] + \sum_{i}\lambda_i\left(\avg{r_i} - \tilde r_i\right)\right].
    \label{eq:F:def}
\end{equation}
Here $\lambda_1, ..., \lambda_n$ respect the stationarity condition.
Then, a critical point occurs when $\mathcal F(\tilde r_1,\cdots ,\tilde r_n)$ is at a saddle point, i.e., the curvature of the free energy is flat along at least one direction, and the resulting symmetry-breaking displacement is related to the associated polarization vector. At the critical point, this condition is reached at the high-symmetry configuration, $\bm{\tilde r}= \bm r^{(0)}$.

For the sake of simplicity, from now on we use mass-rescaled positions, defined as 
\begin{equation}
R_a = \sqrt{m_a}r_a
\label{eq:mass:rescale}
\end{equation}
(see \appendixname~\ref{app:mass:rescale} for details on how any quantity is mass-rescaled).

\eqname~\eqref{eq:F:def} is exact in principle but impractical to solve. The stochastic self-consistent harmonic approximation (SSCHA) restricts the variational landscape to Gaussian density matrices, allowing the explicit minimization of \eqname~\eqref{eq:F:def}\cite{errea_anharmonic_2014,monacelli_stochastic_2021}. The SSCHA name stands for the equivalence between the Gaussian equilibrium density matrix $\hat \rho$ and the auxiliary harmonic Hamiltonian that has $\hat \rho$ as the equilibrium solution. Thus, the SSCHA density matrix is parametrized by the average positions of all atoms $\bm{\mathcal R}$ and the dynamical matrix $\bm D$ of the auxiliary Hamiltonian, which are optimized to minimize the free energy. At the minimum of the free energy, they satisfy the following self-consistent relations\cite{errea_anharmonic_2014,monacelli_stochastic_2021}:
\begin{equation}
    D_{ab} = \avg{\frac{d^2 V}{dR_a dR_b}},\qquad
    \mathcal R_a = \avg{R_a}.
\end{equation}
The auxiliary SCHA dynamical matrix allows us to define the free phonon propagation, with poles at the dressed harmonic frequencies $\omega_\mu$ that diagonalize $\bm D$:
\begin{equation}
    \sum_b e_\mu^bD_{ab} = \omega_\mu^2 e_\mu^a.
\end{equation}

Bianco et al.\cite{bianco_second-order_2017} derived an elegant equation using SSCHA perturbation theory to compute the positional free energy Hessian:
\begin{equation}
    \frac{d^2F}{d\bm R d\bm R} = \bm D + \bm{\dthree}\bm\Lambda \left[
    \mathbbm{1} - \bm{\dfour}\bm\Lambda
    \right]^{-1}\bm{\dthree},
    \label{eq:free:energy:hessian}
\end{equation}
where $\bm\dthree$ and $\bm\dfour$ are the 3, and 4-rank mass-rescaled force constants averaged
over the SSCHA Gaussian density
\begin{equation}
    {\dthree}_{abc} = \avg{ \frac{d^3(V - \mathcal V)}{dR_a dR_b dR_c}},
    \label{eq:d3}
\end{equation}
\begin{equation}
    {\dfour}_{abcd} = \avg{ \frac{d^4(V - \mathcal V)}{dR_a dR_b dR_c dR_d}},
    \label{eq:d4}
\end{equation}
where $\mathcal V(\bm R)$ is the Harmonic potential of the auxiliary hamiltonian:
$$
\mathcal V(\bm R) =\frac{1}{2}\sum_{ab}(R_a - \mathcal R_a)D_{ab}(R_b - \mathcal R_b).
$$
$\bm\Lambda$ is the static limit of the two-phonon free propagator:
\begin{align}
    \Lambda_{abcd} =& -\sum_{\mu\nu}\frac{e_\mu^ae_\nu^b e_\mu^c e_\nu^d}{4\omega_\mu\omega_\nu} \nonumber \\
    &\times
    \begin{cases}
    \displaystyle
        \frac{1 + n_\mu + n_\nu}{\omega_\mu + \omega_\nu} - \frac{n_\mu - n_\nu}{\omega_\mu - \omega_\nu} & \text{for } \omega_\mu\neq\omega_\nu \\[8pt]
        \displaystyle\frac{1 + 2n_\mu}{2\omega_\mu} - \left.\frac{dn}{d\omega}\right|_{\omega = \omega_\mu} & \text{for } \omega_\mu = \omega_\nu.
    \end{cases}
    \label{eq:lambda}
\end{align}
Here, $n_\mu = 1/(e^{\omega_\mu/k_BT} - 1)$ is the Bose-Einstein distribution function, and we assume Hartree atomic units where $\hbar=1$.

\begin{figure}[tbp]
    \centering
    \begin{tikzpicture}[
        phonon/.style={decorate, decoration={snake, amplitude=0.7mm,
            segment length=2.8mm, pre length=0.8mm, post length=0.8mm}, thick},
    ]
    \def\bh{0.25}   

    \node[anchor=east] at (-0.2, 0)
        {$\displaystyle\frac{d^2F}{d\bm{R}^2}\;=\;$};

    \draw[phonon] (0, 0) -- (1.4, 0);
    \node at (1.3, 0.18) {$^{-1}$};
    \node at (0.7, 0.48) {$\bm D$};

    \node at (2.02, 0) {$+$};

    \draw[phonon] (3, \bh) -- (4.30, \bh);          
    \draw[phonon] (3, -\bh) -- (4.30, -\bh);        

    \filldraw[fill=white, draw=black, thick]
        (2.65, 0) -- (3.0, \bh) -- (3.0, -\bh) -- cycle;
    \filldraw[fill=white, draw=black, thick]
        (4.65, 0) -- (4.3, \bh) -- (4.3, -\bh) -- cycle;

    \node at (2.82, -\bh-0.22) {\scriptsize$\bm{\dthree}$};
    \node at (3.65, 0) {$\bm\Lambda$};
    \node at (4.48, -\bh-0.22) {\scriptsize$\bm{\dthree}$};

    \def\yy{-1.4}

    \node at (0.9, \yy) {$+$};

    \draw[phonon] (1.95, \yy+\bh) -- (2.85, \yy+\bh);          
    \draw[phonon] (1.95, \yy-\bh) -- (2.85, \yy-\bh);          
    \draw[phonon] (3.35, \yy+\bh) -- (4.25, \yy+\bh);          
    \draw[phonon] (3.35, \yy-\bh) -- (4.25, \yy-\bh);          

    \filldraw[fill=white, draw=black, thick]
        (1.6, \yy) -- (1.95, \yy+\bh) -- (1.95, \yy-\bh) -- cycle;
    \filldraw[fill=white, draw=black, thick]
        (2.85, \yy+\bh) -- (3.35, \yy+\bh)
     -- (3.35, \yy-\bh) -- (2.85, \yy-\bh) -- cycle;
    \filldraw[fill=white, draw=black, thick]
        (4.6, \yy) -- (4.25, \yy+\bh) -- (4.25, \yy-\bh) -- cycle;

    \node at (1.77, \yy-\bh-0.22) {\scriptsize$\bm{\dthree}$};
    \node at (2.4, \yy) {$\bm\Lambda$};
    \node at (3.1, \yy-\bh-0.22) {\scriptsize$\bm{\dfour}$};
    \node at (3.8, \yy) {$\bm\Lambda$};
    \node at (4.43, \yy-\bh-0.22) {\scriptsize$\bm{\dthree}$};

    \node at (5.4, \yy) {$+\;\cdots$};

    \end{tikzpicture}
    \caption{Diagrammatic representation of the free energy Hessian as an RPA-like resummation.
    Wavy lines are free SCHA phonon propagators: the static one-phonon free propagator is the inverse of the dynamical matrix $\bm D$, while the two-phonon propagator is the $\bm\Lambda$ tensor (\eqname~\ref{eq:lambda}).
    Triangles ($\triangleright$) denote three-phonon vertices $\bm{\dthree}$: each
    triangle has three corners where phonon lines attach (corresponding to the indices to which it can be contracted).
    Squares ($\square$) denote four-phonon vertices $\bm{\dfour}$: each square
    has four corners connecting to the two legs of the two-phonons propagator ${\bm\Lambda}$ on either side. These are the lowest-order diagrams of the geometric expansion of the inversion in \eqname~\eqref{eq:free:energy:hessian}.}
    \label{fig:rpa}
\end{figure}
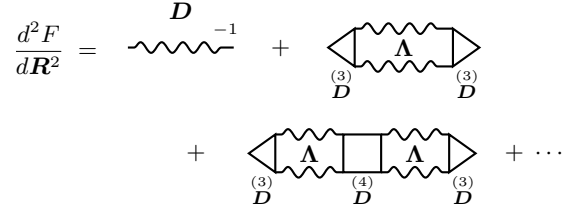

The inversion in \eqname~\eqref{eq:free:energy:hessian} is equivalent to an RPA-like resummation of the ladder diagrams reported in \figname~\ref{fig:rpa}, where $\bm{\dthree}$ and $\bm{\dfour}$ appear as three-phonon and four-phonon scattering vertices, while $\bm D$ is the inverse non-interacting one-phonon propagator.

Truncating the series at the first perturbative term (i.e., neglecting $\bm\dfour$) corresponds to the widely employed \emph{bubble approximation}, where the name takes inspiration from the bubble form of the only surviving self-energy diagram
\begin{equation}
\frac{d^2F}{d\bm R d\bm R} \stackrel{\text{\emph{bubble}}}{\approx} \bm D + \bm{\dthree}\bm\Lambda\bm{\dthree}
.
    \label{eq:free:energy:hessian:bubble}
\end{equation}

While elegant, implementing \eqname~\eqref{eq:free:energy:hessian} as it stands presents two critical problems: (i)~the inversion of a dense $N^2\times N^2$ matrix, and (ii)~the storage of the dense high-rank tensor $\bm{\dfour}$ (\eqname~\ref{eq:d4}). For this reason, the software that implements the free energy Hessian often relies on approximations, like series truncation to lowest order (\eqname~\ref{eq:free:energy:hessian:bubble}) or employing compressed sensing techniques for the storage of high-rank force constant tensors\cite{phono3py,tadano_anharmonic_2014,eriksson_hiphive_2019-1}.

Here, we devise a different approach that completely avoids materializing either $\bm\dthree$ or $\bm{\dfour}$ while maintaining the exact equivalence to the full \eqname~\eqref{eq:free:energy:hessian}. The key insight is to reformulate \eqname~\eqref{eq:free:energy:hessian} as a linear system, and solve it iteratively by applying only sparse matrix-vector products. 

\section{Positional free energy Hessian as a sparse linear system}
\label{sec:linear:system}
Let us define a vector space of possible perturbations of the equilibrium Gaussian density matrix to static external perturbations as 
\begin{equation}
    \ket\psi = \begin{pmatrix}
    \bm {\mathcal R^{(1)}} \\ 
    \bm {\Upsilon^{(1)}}
    \end{pmatrix},
\end{equation}
where the space is composed in two sectors: $\bm{\mathcal R^{(1)}}$ and $\bm{\Upsilon^{(1)}}$. $\bm{\mathcal R^{(1)}}$ has dimension $3N$ and represents the perturbation on the average positions of atoms ($N$ atoms and 3 Cartesian directions). $\bm{\Upsilon^{(1)}}$ has dimension $3N(3N+1)/2$, and represents the perturbation of the Gaussian covariance matrix (thus a symmetric $(3N)\times (3N)$ matrix). 
The positional free energy Hessian can be computed by solving a linear system in this vector space as
\begin{equation}
\calL \ket{\phi_k} = \ket{\bm{\delta_k}}
\label{eq:linear:system}
\end{equation}
where
\begin{equation}
    {\calL} = \begin{pmatrix}
        \bm{D}&
        \bm{\dthree} \\
        \bm{\dthree}&
        \bm{\dfour} -\bm{\Lambda}^{-1}
    \end{pmatrix},
    \quad
    \ket{\bm{\delta_k}} = \begin{pmatrix}
        \bm{\delta_k}\\0
    \end{pmatrix}.
    \label{eq:L:matrix}
\end{equation}
The $\calL$ is an operator in this vector space determining how the density matrix propagates. Lastly, the $\bm{\delta_k}$ is a $3N$ vector whose all elements are 0 except the $k$-th index, which is 1.
In the SSCHA language, the $\bm {\mathcal R^{(1)}}$ and $\bm{\Upsilon^{(1)}}$ blocks correspond, respectively, to the propagation of one and two phonons. 

The free energy Hessian can be obtained from the inverse of $\calL$ by solving the linear system in \eqname~\eqref{eq:linear:system}: let $\bm X^k$ be the one-phonon block of the solution of \eqname~\eqref{eq:linear:system}, then the $k$-th column of the inverse of the positional free energy Hessian is $\bm X^k$. 
\begin{equation}
    \ket{\phi_k} = \begin{pmatrix}
        \bm {X^k} \\ 
        \bm {Y^k}
    \end{pmatrix}
\end{equation}

Iterating over $k$, we can reconstruct the full inverse of the free energy Hessian (i.e., the static susceptibility) and, therefore, the Hessian itself:
\begin{equation}
    \bm{\calG} = \begin{pmatrix}
        X^{1}_1 & X^{2}_1 & \cdots & X^{3N}_1 \\
        X^{1}_2 & \ddots & \cdots  & X^{3N}_2 \\
        \vdots & \vdots & \ddots & \vdots \\
        X_{3N}^{1} & X^{2}_{3N} & \cdots & X_{3N}^{3N}
    \end{pmatrix},
    \label{eq:static:G}
\end{equation}
\begin{equation}
    \frac{d^2F}{dR_adR_b} = \left({\bm \calG}^{-1}\right)_{ab}.
    \label{eq:hessian:inverse}
\end{equation}

The proof of the equivalence between \eqname~\eqref{eq:hessian:inverse} and \eqname~\eqref{eq:free:energy:hessian} is reported in detail in \appendixname~\ref{app:equivalence:proof}. 

The linear system in \eqname~\eqref{eq:linear:system} can be solved efficiently with the GMRES algorithm, once we have a practical recipe to apply the $\calL$ operator on a generic $\ket{\psi}$ vector. The GMRES algorithm should be chosen over conjugate gradient approaches because the $\calL$ matrix is not positive definite when the free-energy Hessian is a saddle point. The details of the numerical solver are discussed in \appendixname~\ref{app:linear:system}.

\subsection{Efficient implementation}
\label{subsec:Efficient:implementation}
Recasting the free-energy Hessian calculation as a linear system allows us to exploit efficient iterative algorithms developed for this purpose (see \appendixname~\ref{app:linear:system}). However, inverting \eqname~\eqref{eq:linear:system} is convenient only if the operator $\bm\calL$ is sparse, which is not the case for a general system in \eqname~\eqref{eq:L:matrix}, where all phonons interact. 
However, here we prove that applying the operator $\calL$ to a generic vector $\ket{\psi}$ can be achieved in $O(N)$ operations, i.e., without materializing $\bm{\dfour}$ and $\bm{\dthree}$.
A similar technique was already introduced for the dynamical Lanczos algorithm of TD-SCHA\cite{monacelli_time-dependent_2021}.

We decompose $\calL = \calL^{(\text{har})} + \calL^{(\text{anh})}$, where
\begin{equation}
    \calL^{(\text{har})} = \begin{pmatrix}
        \bm{D} & 0 \\
        0 & -\bm{\Lambda}^{-1}
    \end{pmatrix}, \quad
    \calL^{(\text{anh})} = \begin{pmatrix}
        0 & \bm{\dthree} \\
        \bm{\dthree} & \bm{\dfour}
    \end{pmatrix}.
    \label{eq:L:decomposition}
\end{equation}
The harmonic part $\calL^{(\text{har})}$ is diagonal in the phonon polarization basis. The anharmonic part $\calL^{(\text{anh})}$ is the expensive contribution.

The application of $\calL^\text{(anh)}$ on the two blocks of a generic $\ket \psi$ vector can be written as
$$
\calL^\text{(anh)}\ket\psi = \begin{pmatrix}
    \left[\cal L^\text{(anh)} \ket\psi\right]_{\bm {\mathcal R^{(1)}}} \\ 
    \left[\cal L^\text{(anh)} \ket\psi\right]_{\bm {\Upsilon^{(1)}}} 
\end{pmatrix}.
$$
Each component of this vector is explicitly expressed as
\begin{equation}
    \left[\mathcal L^\text{(anh)}\ket\psi\right]_{{\mathcal R^{(1)}}_a} = \sum_{bc}\dthree_{abc}\Upsilon^{(1)}_{bc},
    \label{eq:R1:sector}
\end{equation}
\begin{equation}
    \left[\mathcal L^\text{(anh)}\ket\psi\right]_{{\Upsilon^{(1)}}_{ab}} = \sum_{c}\dthree_{abc}\mathcal R^{(1)}_{c} + \sum_{cd}\dfour_{abcd}\Upsilon^{(1)}_{cd}.
    \label{eq:Y1:sector}
\end{equation}
This expression is still computationally extremely expensive.

However, the SSCHA defines the high-rank tensors $\bm \dthree$ and $\bm\dfour$ as stochastic averages over displacements and forces (see ref.~\cite{bianco_second-order_2017}).
\begin{equation}
    \dthree_{abc} = -\sum_{hk}\Upsilon_{ah}\Upsilon_{bk}\avg{u_h u_k f_c}
    \label{eq:d3:new}
\end{equation}
\begin{equation}
    \dfour_{abcd} = -\sum_{hkl}\Upsilon_{ah}\Upsilon_{bk}\Upsilon_{cl}\avg{u_h u_k u_lf_d}
    \label{eq:d4:new}
\end{equation}
where $\bm u$ is the centroid displacement for each configuration in the ensemble $(\bm R - \bm {\mathcal R})$ and $\bm f$ is the BO force subtracted from the auxiliary harmonic force: 
$$
u_a = (R_a - \mathcal R_a), \qquad
f_a = -\frac{dV}{dR_a} +\sum_c D_{ac}u_c.
$$
The average $\avg{\cdot}$ is performed over the equilibrium SSCHA Gaussian density matrix.
Note that both $\bm f$ and $\bm u$ are mass-rescaled quantities (see \appendixname~\ref{app:mass:rescale}). Analogously, $\bm\Upsilon$ is the (mass-rescaled) inverse covariance matrix:
\begin{equation}
    \Upsilon_{ab} = \sum_\mu e_\mu^a e_\mu^b \frac{2\omega_\mu}{2n_\mu + 1}.
\end{equation}
To compact the notation of the high-rank tensor in \eqname~\eqref{eq:d3:new} and \eqref{eq:d4:new}, we can define a new vector $\bm v$ as the product between $\bm\Upsilon$ and the displacement $\bm u$
$$
v_a = \sum_h\Upsilon_{ah}u_h.
$$

The advantage of \eqname~\eqref{eq:d3:new} and \eqname~\eqref{eq:d4:new} is that the high-rank tensors can be written as averages of the outer product between vectors. Replacing \eqname~\eqref{eq:d3:new} and \eqname~\eqref{eq:d4:new} inside \eqname~\eqref{eq:R1:sector} and \eqname~\eqref{eq:Y1:sector}, we get
\begin{equation}
    \left[\mathcal L^\text{(anh)}\ket\psi\right]_{{\mathcal R^{(1)}}_a} = -\avg{
    v_a \sum_{bc}v_bf_c \Upsilon^{(1)}_{bc}},
    \label{eq:R1:sector:fast}
\end{equation}
\begin{align}
    \left[\mathcal L^\text{(anh)}\ket\psi\right]_{{\Upsilon^{(1)}}_{ab}}& = 
    -\avg{v_a v_b \sum_{c}f_c\mathcal R^{(1)}_{c}}  \nonumber \\ 
    &-\avg{v_a v_b\sum_{cd}v_cf_d\Upsilon^{(1)}_{cd}},
    \label{eq:Y1:sector:fast}
\end{align}

By introducing a configuration dependent weight $w[\psi](\bm R)$, the application of $\calL^\text{(anh)}$ becomes 

\begin{align}
    \left[\mathcal L^\text{(anh)}\ket\psi\right]_{{{\mathcal R}^{(1)}}_{a}} & = \avg{
        w[\psi]\,
        f_a
        }\nonumber \\
  \left[\mathcal L^\text{(anh)}\ket\psi\right]_{{\Upsilon^{(1)}}_{ab}} &= -\avg{
        w[\psi]\,
        \frac{d^2(V-\mathcal V)}
        {d\tilde R_a\, d\tilde R_b}
        }
    \label{eq:L:anharmonic:application}
\end{align}

\begin{equation}
    w[\psi](\bm R) = \sum_{a} \mathcal R^{(1)}_a \frac{\partial\ln\rho}{\partial \mathcal R_a} +2 \sum_{abhk}\Upsilon^{(1)}_{ab}\Upsilon_{ah}\Upsilon_{bk}\frac{\partial\ln\rho}{\partial\Upsilon_{hk}}
    \label{eq:weight}.
\end{equation}

We recall that $\rho$ is the Gaussian equilibrium density matrix from the SSCHA Hamiltonian, i.e.
\begin{equation}
    \rho(\bm R) = \sqrt\frac{\det\Upsilon}{(2\pi)^{3N}} \exp\left[-\frac 12 \sum_{ab}(R_a - \mathcal R_a)\Upsilon_{ab}(R_b - \mathcal R_b)\right]
    \label{eq:rho}
\end{equation}

\eqname~\eqref{eq:L:anharmonic:application} has the great advantage of never materializing the high-rank force-constants $\bm\dthree$ and $\bm\dfour$, providing a much faster implicit way to define them; each matrix-vector product therefore requires only $\mathcal{O}(N_\text{modes}^2)$ operations per configuration, compared to $\mathcal{O}(N_\text{modes}^4)$ for explicit construction of $\bm{\dfour}$. This trick was first introduced in the TD-SCHA Lanczos algorithm from ref.\cite{monacelli_time-dependent_2021}. 
The weight $w[\psi]$ can be decomposed into the contributions given by the centroid block $\bm{{\mathcal R}^{(1)}}$ and the two-phonons block $\bm{\Upsilon^{(1)}}$: 
$$
w[\psi](\bm R) =  w[\bm{\mathcal R^{(1)}}](\bm R) + w[\bm {\Upsilon^{(1)}}](\bm R)
$$
\begin{equation}
    w[\bm{\mathcal R^{(1)}}](\bm R) = -\sum_{a} \mathcal R_a^{(1)}v_a
\end{equation}
\begin{equation}
    w[\bm {\Upsilon^{(1)}}](\bm R) = -\sum_{ab}\Upsilon^{(1)}_{ab}v_a v_b
    \label{eq:w:Y}
\end{equation}
where the derivative of the normalization factor in \eqname~\eqref{eq:rho} can be omitted, as it leads to a weight independent of the atomic position that averages to zero in the equilibrium SSCHA result.
In particular, the $\avg{w[\bm {{\mathcal R}^{(1)}}] f_a}$ is zero in the SSCHA equilibrium and can be omitted, resulting in

\begin{align}
    \left[\mathcal L^\text{(anh)}\ket\psi\right]_{{{\mathcal R}^{(1)}}_{a}} & = \avg{
        w[{\bm {\Upsilon^{(1)}}}]\,
        f_a
        }\nonumber \\
  \left[\mathcal L^\text{(anh)}\ket\psi\right]_{{\Upsilon^{(1)}}_{ab}} &= -\avg{
        \left(w[\bm{{\mathcal R}^{(1)}}]
        + w[{\bm {\Upsilon^{(1)}}}]
        \right) \,
        \frac{d^2(V-\mathcal V)}
        {d R_a\, d R_b}
        }
    \label{eq:quick:L}
\end{align}
Indeed, the permutation symmetry of the $\bm\dthree$ and $\bm\dfour$ indices can be exploited by averaging the results obtained by swapping the force $f$ with the $v$ variables.

Applying \eqname~\eqref{eq:quick:L} is still very expensive, as the $\ket{\bm\delta_k}$ perturbation breaks translational symmetry.
Therefore, translational symmetries inside the supercell must be imposed \emph{a posteriori} by averaging the result among symmetry-equivalent perturbations, leading to a still unfavorable $\mathcal{O}(N^3)$ scaling: $N$ per translation times $N^2$ for the $w[{\bm {\Upsilon^{(1)}}}]$ and the $d(V-\mathcal V)/dR_adR_b$. 

The promising linear scaling is achieved in \secname~\ref{sec:qspace}, where we reformulate \eqname~\eqref{eq:quick:L} in reciprocal space. There, momentum conservation reduces the computational cost by a factor of $N_q^2$, where $N_q$ is the q-space mesh, thus leading to linear scaling ($N = N_qN_\text{uc}$, where $N_\text{uc}$ is the number of atoms in the primitive cell and fixed).

\subsection{TD-SCHA in momentum basis}\label{sec:qspace}

Linear scaling with supercell size can be achieved thanks to momentum conservation. This property can be easily enforced on the vector $\ket{\psi}$ when the external perturbation is monochromatic, therefore avoiding averaging over all translational replicas of \eqname~\eqref{eq:quick:L}.
Moreover, the momentum basis also sparsifies $\calL^\text{(anh)}$ thanks to the Bloch theorem.

Let us start with the definition of a vector in momentum space. Let $v_\kappa(\bm R)$ be a real-space vector, where $\kappa$ is the atom-cartesian index in the primitive cell, while $\bm R$ is a lattice vector indicating on which supercell the $\kappa$ atom is. $v_\kappa(\bm R)$ transforms in $v_\kappa(\bm q)$ as
\begin{equation}
    v_\kappa(\bm q) = \frac{1}{\sqrt{N_q}}\sum_{\bm R} e^{-2\pi i \bm R\cdot\bm q}v_{\kappa}(\bm R).
    \label{eq:fourier}
\end{equation}
This phase convention also defines the form of the dynamical matrix in $\bm q$ space:
\begin{equation}
    D_{\kappa\kappa'}(\bm q) = \sum_{\bm R} e^{- 2\pi i\bm R\cdot\bm q}D_{\kappa,\,\kappa'}(\bm R),
    \label{eq:dyn:fourier}
\end{equation}
where $D_{\kappa \kappa'}(\bm R)$ is the force-constant matrix between atom \(\kappa\) in the reference primitive cell and atom \(\kappa'\) in another primitive cell. The vector \(\bm R\) is the lattice vector connecting the origins of the two primitive cells. In $\bm q$-space, the dynamical matrix has eigenvalues $\omega_\mu^2(\bm q)$ and eigenvectors $e_\mu^\kappa(\bm q)$.

The key advantage of the $\bm q$-space formulation arises when we use a monochromatic external perturbation at a given $\bm q$ vector; then the response must also be at the same $\bm q$ vector (modulo a reciprocal lattice vector $\bm G$). This means that the perturbed centroids oscillate only at momentum $\bm q$, reducing the degrees of freedom in the $\bm {\mathcal R^{(1)}}$ sector from the atoms in the supercell ($3N_\text{sc}$) to the atoms on the primitive cell ($3N_\text{uc}$). Even better for the covariance matrix perturbation $\bm {\Upsilon^{(1)}}$, which, when transformed in Fourier space, becomes:
\begin{equation}
    \Upsilon^{(1)}_{ab}(\bm q_1, \bm q_2) = \frac{1}{N_q}\sum_{\bm R_1,\bm R_2}e^{-2\pi i \left(\bm q_2\cdot \bm R_2 + \bm q_1\cdot\bm R_1\right)} {\Upsilon^{(1)}}_{ab}(\bm {R_1}, \bm {R_2}).
\end{equation}

The crucial observation is that $\Upsilon^{(1)}(\bm q_1, \bm q_2)$ is excited from the centroid sector $\bm R^{(1)}(\bm q)$ only through the the three-phonon vertex $\bm{\dthree}$, see \eqname~\eqref{eq:Y1:sector}. In $\bm q$ space, $\bm\dthree$ must conserve the momentum so that the sum of the incoming $\bm q$ (coming from $\bm{\mathcal R^{(1)}}$) must be equal to the sum of the outgoing momenta $\bm q_1 + \bm q_2$ (modulo a reciprocal lattice vector $\bm G$)
\begin{equation}
    \bm q_1 + \bm q_2 = \bm q + \bm G.
\end{equation}

\begin{figure}[tbp]
    \centering
    \begin{tikzpicture}[
        phonon/.style={decorate, decoration={snake, amplitude=0.7mm,
            segment length=2.8mm, pre length=0.8mm, post length=0.8mm}, thick},
    ]
    \coordinate (L)  at (-0.5, 0);
    \coordinate (TR) at ( 0.3, 0.45);
    \coordinate (BR) at ( 0.3,-0.45);

    \draw[phonon] (-2.6, 0) -- (L);
    \draw[phonon] (TR) -- ( 2.2, 1.15);
    \draw[phonon] (BR) -- ( 2.2,-1.15);

    \filldraw[fill=white, draw=black, thick] (L) -- (TR) -- (BR) -- cycle;

    \node at (0.05, 0.03) {\small$\bm{\dthree}$};
    \node at (-1.5, 0.3)  {$\bm{q}$};
    \node at ( 1.5, 1.2) {$\bm{q}_1$};
    \node at ( 1.5,-1.2) {$\bm{q}_2$};

    \node at (0, -1.7) {$\bm{q}_1 + \bm{q}_2 = \bm{q} + \bm{G}$};
    \end{tikzpicture}
    \caption{Three-phonon scattering vertex $\bm{\dthree}$, represented as a
    triangle whose three corners connect to phonon lines.
    A perturbation at wave vector $\bm{q}$ (left vertex) decays into two
    phonons with momenta $\bm{q}_1$ and $\bm{q}_2$ (right vertices)
    satisfying momentum conservation $\bm{q}_1 + \bm{q}_2 = \bm{q} + \bm{G}$.}
    \label{fig:decay}
\end{figure}
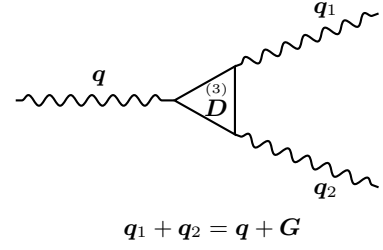

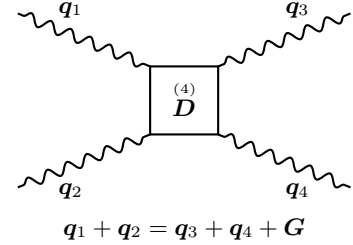
\begin{figure}[tbp]
    \centering
    \begin{tikzpicture}[
        phonon/.style={decorate, decoration={snake, amplitude=0.7mm,
            segment length=2.8mm, pre length=0.8mm, post length=0.8mm}, thick},
    ]
    \coordinate (TL) at (-0.45, 0.45);
    \coordinate (BL) at (-0.45,-0.45);
    \coordinate (TR) at ( 0.45, 0.45);
    \coordinate (BR) at ( 0.45,-0.45);

    \draw[phonon] (-2.2, 1.15) -- (TL);
    \draw[phonon] (-2.2,-1.15) -- (BL);
    \draw[phonon] (TR) -- ( 2.2, 1.15);
    \draw[phonon] (BR) -- ( 2.2,-1.15);

    \filldraw[fill=white, draw=black, thick] (TL) -- (TR) -- (BR) -- (BL) -- cycle;

    \node at (0, 0.03) {\small$\bm{\dfour}$};
    \node at (-1.5, 1.2) {$\bm{q}_1$};
    \node at (-1.5,-1.2) {$\bm{q}_2$};
    \node at ( 1.5, 1.2) {$\bm{q}_3$};
    \node at ( 1.5,-1.2) {$\bm{q}_4$};

    \node at (0, -1.7) {$\bm{q}_1 + \bm{q}_2 = \bm{q}_3 + \bm{q}_4 + \bm{G}$};
    \end{tikzpicture}
    \caption{Four-phonon scattering vertex $\bm{\dfour}$, represented as a
    square whose four corners connect to phonon lines.
    Two incoming phonons with momenta $\bm{q}_1$ and $\bm{q}_2$ (left corners)
    scatter into two outgoing phonons with momenta $\bm{q}_3$ and $\bm{q}_4$
    (right corners), satisfying momentum conservation
    $\bm{q}_1 + \bm{q}_2 = \bm{q}_3 + \bm{q}_4 + \bm{G}$.}
    \label{fig:decay4}
\end{figure}

The application of $\mathcal L^\text{(anh)}$ on the covariance matrix sector reads
\begin{align}
    \left[\mathcal L^\text{(anh)}\ket\psi\right]&{}_{\Upsilon^{(1)}_{ab}(\bm q_1, \bm q_2)} = \sum_{c,\bm q_3}\dthree_{abc}(\bm q_1, \bm q_2, -\bm q_3)\mathcal R^{(1)}_c(\bm q_3)  \nonumber \\ 
    & +\sum_{\substack{cd\\\bm q_3, \bm q_4}} \dfour_{abcd}(\bm q_1, \bm q_2, -\bm q_3,  -\bm q_4)\Upsilon^{(1)}_{cd}(\bm q_3, \bm q_4).
    \label{eq:lanh:qtot}
\end{align}
The two processes can be depicted in diagrams of \figurename~\ref{fig:decay} and \ref{fig:decay4}, where $\bm{\mathcal R^{(1)}}(\bm q)$ represents a one-phonon propagation while $\bm{\Upsilon^{(1)}}(\bm {q_1},\bm {q_2})$ represents the two-phonons propagation.
If the original perturbation is monochromatic in $\bm q$, and the centroid is constrained only on $\bm{\mathcal R^{(1)}}(\bm q)$, and imposing the momentum conservation rule inside the $\bm\dthree$ and $\bm\dfour$, we immediately see that $\bm \Upsilon^{(1)}$ is constrained only to be defined at $\bm q_1, \bm q - \bm q_1$ (for any possible $\bm q_1$)
\begin{align}
    \bigg[\mathcal L^\text{(anh)}&\ket\psi\bigg]{}_{\Upsilon^{(1)}_{ab}(\bm q_1, \bm q - \bm q_1)} = \sum_{c}\dthree_{abc}(\bm q_1, \bm q - \bm q_1, -\bm q)\mathcal R^{(1)}_c(\bm q)  \nonumber \\ 
    & +\sum_{\substack{cd\\\bm q_2}} \dfour_{abcd}(\bm q_1, \bm q - \bm q_1, -\bm q_2, \bm q_2 - \bm q)\Upsilon^{(1)}_{cd}(\bm q_2, \bm q - \bm q_2).
    \label{eq:lanh:qcons}
\end{align}
The reduction in both the degrees of freedom of $\bm {\mathcal R^{(1)}}$ and ${\bm \Upsilon^{(1)}}$, as well as the drop of one $\bm {q_3}$ summation between \eqname~\eqref{eq:lanh:qtot} to \eqref{eq:lanh:qcons} is the key advantage of using $\bm q$-space basis. Indeed, this reduction is only possible if we consider monochromatic perturbations, which is, however, quite general and true for most physically relevant observables (e.g., Raman and IR spectroscopy probe the $\bm q\to 0$ response function, inelastic neutron and X-ray scattering measure the dynamical structure factor resolved at specific $\bm q$ momenta).
Indeed, since we require perturbing $\ket{\bm{\delta_k}}$ on all supercell atoms for computing the free energy Hessian in \eqname~\eqref{eq:linear:system}, we can change basis without loss of generality and pick $\ket{\bm{\tilde{\delta_k}}(\bm q)}$ as monochromatic perturbations.

The momentum conservation, illustrated in \figname~\ref{fig:decay} and \ref{fig:decay4}, implies that the static Liouvillian $\calL$ is block-diagonal in $\bm q$: for each perturbation wave vector $\bm q$, the $\bm{\mathcal R^{(1)}}$-sector is always only evaluated at the perturbation $\bm q$, while the $\bm{\Upsilon^{(1)}}$-sector contains two-phonon blocks labeled by the $N_q$ momentum-conserving pairs $(\bm q_1, \bm q_2)$ with $\bm q_1 + \bm q_2 = \bm q + \bm G$,
\begin{equation}
    \ket{\psi_{\bm q}} = \begin{pmatrix}
        \bm {\mathcal R^{(1)}}(\bm q) \\ 
        \bm {\Upsilon^{(1)}}(\bm {q_1}, \bm q - \bm {q_1})
    \end{pmatrix}.
\end{equation}
As a consequence, fixing $\bm q$, the perturbation $\bm{\Upsilon^{(1)}}$ has the same degrees of freedom as a dynamical matrix, i.e., $(3N_\text{uc})^2N_q$.

Fourier transforming the stochastic implementation presented in \eqname~\ref{eq:R1:sector:fast} and \ref{eq:Y1:sector:fast}, we get
\begin{equation}
    \left[\mathcal L^{\text{(anh)}} \ket {\psi_{\bm q}}\right]_{{\mathcal R^{(1)}_a(\bm q)}} = 
    \avg{w[\bm {\Upsilon^{(1)}_{q}}] f_a^{(i)}(\bm q)},
\end{equation}
\begin{align}
    \left[\mathcal L^{\text{(anh)}}\ket {\psi_{\bm q}}\right]&_{{\Upsilon^{(1)}_{ab}(\bm q_1, \bm q - \bm {q_1})}} = 
    \bigg< \left(w[\bm {\mathcal R^{(1)}_q}] + w[\bm {\Upsilon^{(1)}_q}]\right) \nonumber \\
    & \sum_{h}\Upsilon_{ah}(\bm q_1)u_h^{(i)}(\bm q_1)f_b^{(i)}(\bm q - \bm {q_1})\bigg>_\rho,
    \label{eq:Lapply:q:Y}
\end{align}
where the weight can also be computed in Fourier space by transforming \eqname~\eqref{eq:weight}
\begin{equation}
    w_i[\bm{\mathcal R^{(1)}_q}] = -\sum_{ah} \mathcal R_a^{(1)}(-\bm q)\Upsilon_{ah}(\bm q)u_h^{(i)}(\bm q),
\end{equation}
\begin{align}
    w_i[\bm {\Upsilon^{(1)}}] = -\sum_{\substack{abhk\\ \bm {q_2}}}\Upsilon^{(1)}_{ab}(-\bm q_2, &\bm {q_2} - \bm q)\Upsilon_{ah}(\bm {q_2})u^{(i)}_h(\bm {q_2}) \nonumber \\ 
    &  \Upsilon_{bk}(\bm q - \bm{q_2})u^{(i)}_k(\bm q - \bm {q_2}).
    \label{eq:w:Y:q}
\end{align}

\begin{figure}[tbp]
    \centering
    \resizebox{\columnwidth}{!}{%
    \begin{tikzpicture}[
        phonon/.style={decorate, decoration={snake, amplitude=0.7mm,
            segment length=2.8mm, pre length=0.8mm, post length=0.8mm}, thick},
    ]
    \coordinate (LA) at (-4.0, 0.00);
    \coordinate (LT) at (-3.3, 0.50);
    \coordinate (LB) at (-3.3,-0.50);
    \coordinate (STL) at (-0.5, 0.50);
    \coordinate (SBL) at (-0.5,-0.50);
    \coordinate (STR) at ( 0.5, 0.50);
    \coordinate (SBR) at ( 0.5,-0.50);
    \coordinate (RA) at ( 4.0, 0.00);
    \coordinate (RT) at ( 3.3, 0.50);
    \coordinate (RB) at ( 3.3,-0.50);

    \draw[phonon] (-5.0, 0) -- (LA) node[midway, above=1.5mm] {$\bm{q}$};
    \draw[phonon] (RA) -- ( 5.0, 0) node[midway, above=1.5mm] {$\bm{q}$};

    \draw[phonon] (LT) to[out=40,  in=140]  node[midway, above=1.5mm] {$\bm{q}_1$} (STL);
    \draw[phonon] (LB) to[out=-40, in=-140] node[midway, below=1.5mm] {$\bm{q}_2$} (SBL);
    \draw[phonon] (STR) to[out=40,  in=140]  node[midway, above=1.5mm] {$\bm{q}_3$} (RT);
    \draw[phonon] (SBR) to[out=-40, in=-140] node[midway, below=1.5mm] {$\bm{q}_4$} (RB);

    \filldraw[fill=white, draw=black, thick] (LA) -- (LT) -- (LB) -- cycle;
    \filldraw[fill=white, draw=black, thick] (RA) -- (RT) -- (RB) -- cycle;
    \filldraw[fill=white, draw=black, thick] (STL) -- (STR) -- (SBR) -- (SBL) -- cycle;

    \node at (-3.53, 0) {\small$\bm{\dthree}$};
    \node at ( 3.53, 0) {\small$\bm{\dthree}$};
    \node at ( 0.00, 0) {\small$\bm{\dfour}$};

    \node at (-1.9, 0) {$\bm{\Lambda}$};
    \node at ( 1.9, 0) {$\bm{\Lambda}$};

    \node at (0, -2.0)
        {$\bm{q}_1 + \bm{q}_2 = \bm{q}_3 + \bm{q}_4 = \bm{q} \quad (\mathrm{mod}\ \bm{G})$};
    \end{tikzpicture}}
    \caption{Fourth-order contribution
    $\bm{\dthree}\bm{\Lambda}\bm{\dfour}\bm{\Lambda}\bm{\dthree}$ to the free
    energy Hessian in momentum space. An incoming phonon of quasi-momentum
    $\bm{q}$ splits at a cubic vertex into the pair $(\bm{q}_1,\bm{q}_2)$,
    which propagates through $\bm{\Lambda}$, is scattered by the quartic
    vertex $\bm{\dfour}$ into the pair $(\bm{q}_3,\bm{q}_4)$, and recombines
    at the second cubic vertex into an outgoing phonon of the same
    quasi-momentum. Quasi-momentum is conserved at each vertex up to a
    reciprocal lattice vector $\bm{G}$.}
    \label{fig:fourth-order}
\end{figure}
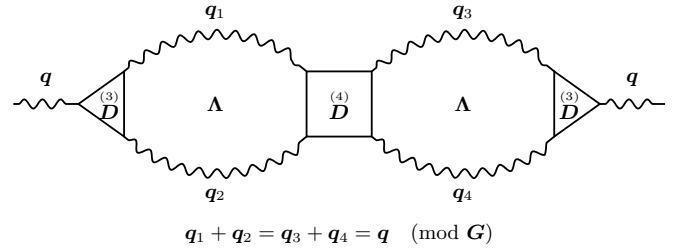

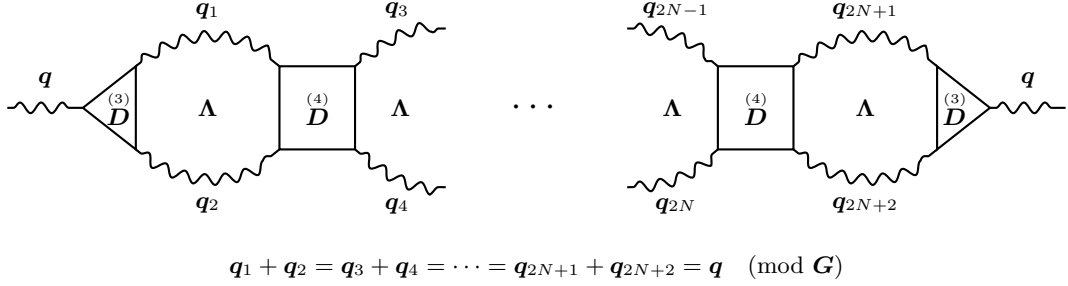
\begin{figure*}[tbp]
    \centering
    \begin{tikzpicture}[
        phonon/.style={decorate, decoration={snake, amplitude=0.7mm,
            segment length=2.8mm, pre length=0.8mm, post length=0.8mm}, thick},
    ]
    \coordinate (LA) at (-6.0, 0.00);
    \coordinate (LT) at (-5.3, 0.55);
    \coordinate (LB) at (-5.3,-0.55);
    \coordinate (S1TL) at (-3.4, 0.55);  \coordinate (S1BL) at (-3.4,-0.55);
    \coordinate (S1TR) at (-2.4, 0.55);  \coordinate (S1BR) at (-2.4,-0.55);
    \coordinate (S2TL) at ( 2.4, 0.55);  \coordinate (S2BL) at ( 2.4,-0.55);
    \coordinate (S2TR) at ( 3.4, 0.55);  \coordinate (S2BR) at ( 3.4,-0.55);
    \coordinate (RT) at ( 5.3, 0.55);
    \coordinate (RB) at ( 5.3,-0.55);
    \coordinate (RA) at ( 6.0, 0.00);

    \draw[phonon] (-7.0, 0) -- (LA) node[midway, above=1.5mm] {$\bm{q}$};
    \draw[phonon] (RA) -- ( 7.0, 0) node[midway, above=1.5mm] {$\bm{q}$};

    \draw[phonon] (LT) to[out=40,  in=140]  node[midway, above=1.5mm] {$\bm{q}_1$} (S1TL);
    \draw[phonon] (LB) to[out=-40, in=-140] node[midway, below=1.5mm] {$\bm{q}_2$} (S1BL);
    \draw[phonon] (S1TR) to[out=40,  in=180] node[midway, above=1.5mm] {$\bm{q}_3$} (-1.2, 1.05);
    \draw[phonon] (S1BR) to[out=-40, in=180] node[midway, below=1.5mm] {$\bm{q}_4$} (-1.2,-1.05);
    \draw[phonon] ( 1.2, 1.05) to[out=0, in=140]
        node[midway, above=1.5mm] {$\bm{q}_{2N-1}$} (S2TL);
    \draw[phonon] ( 1.2,-1.05) to[out=0, in=-140]
        node[midway, below=1.5mm] {$\bm{q}_{2N}$} (S2BL);
    \draw[phonon] (S2TR) to[out=40,  in=140]
        node[midway, above=1.5mm] {$\bm{q}_{2N+1}$} (RT);
    \draw[phonon] (S2BR) to[out=-40, in=-140]
        node[midway, below=1.5mm] {$\bm{q}_{2N+2}$} (RB);

    \filldraw[fill=white, draw=black, thick] (LA) -- (LT) -- (LB) -- cycle;
    \filldraw[fill=white, draw=black, thick] (RA) -- (RT) -- (RB) -- cycle;
    \filldraw[fill=white, draw=black, thick] (S1TL) -- (S1TR) -- (S1BR) -- (S1BL) -- cycle;
    \filldraw[fill=white, draw=black, thick] (S2TL) -- (S2TR) -- (S2BR) -- (S2BL) -- cycle;

    \node at (-5.53, 0) {\small$\bm{\dthree}$};
    \node at ( 5.53, 0) {\small$\bm{\dthree}$};
    \node at (-2.90, 0) {\small$\bm{\dfour}$};
    \node at ( 2.90, 0) {\small$\bm{\dfour}$};

    \node at (-4.35, 0) {$\bm{\Lambda}$};
    \node at (-1.80, 0) {$\bm{\Lambda}$};
    \node at ( 1.80, 0) {$\bm{\Lambda}$};
    \node at ( 4.35, 0) {$\bm{\Lambda}$};

    \node at (0, 0) {\Large$\cdots$};

    \node at (0, -2.1) {$\bm{q}_1 + \bm{q}_2 = \bm{q}_3 + \bm{q}_4
        = \cdots = \bm{q}_{2N+1} + \bm{q}_{2N+2}
        = \bm{q} \quad (\mathrm{mod}\ \bm{G})$};
    \end{tikzpicture}
    \caption{Generic diagram in the Taylor expansion in powers of $\dfour$ of the inverse in \eqname~\eqref{eq:free:energy:hessian}. The term represents a contribution of type
    $\bm{\dthree}\,\bm{\Lambda}\,\bm{\dfour}\,\bm{\Lambda}\cdots
    \bm{\Lambda}\,\bm{\dfour}\,\bm{\Lambda}\,\bm{\dthree}$ to the free energy
    Hessian in eq. \eqref{eq:free:energy:hessian} in momentum space, containing $N$ quartic vertices and $N+1$
    two-phonon propagators $\bm{\Lambda}$. The incoming phonon of
    quasi-momentum $\bm{q}$ splits at the first cubic vertex into a phonon
    pair which is repeatedly scattered by the quartic vertices before
    recombining at the second cubic vertex. Each $\bm{\Lambda}$ carries a
    pair $(\bm{q}_{2k-1}, \bm{q}_{2k})$ whose total quasi-momentum equals
    $\bm{q}$ up to a reciprocal lattice vector, so that the resummation of
    all such terms is exactly the geometric series solved by the iterative
    algorithm.}
    \label{fig:generic-order}
\end{figure*}

The formulation in $\bm q$ space has enormous advantages: (i) all the sums of atomic indices run on the primitive cell, making the application of $\mathcal L$ scale as $O(N_\text{uc}^2N_q)$, and (ii) the momentum conservation is automatically enforced, so there is no need to average over all lattice translations. The cost of applying $\calL$, taking also symmetries into account, scales as $O(N_\text{uc}^2 N_q N_G)$, against an $O(N_\text{uc}^2 N_q^3N_G)$ for the same implementation in real space ($N_G$ is the number of symmetries in the point group).
Notably, this is a better scaling even than evaluating the perturbative diagrams individually, since any $\bm\dfour$ vertex introduces an additional summation over $\bm q$, as depicted in the lowest-order $\bm\dfour$ RPA diagram (\figurename~\ref{fig:fourth-order}) already carries an intrinsic $N_q^2$ momentum sum. Indeed, evaluating the RPA series perturbatively introduces an additional power to the $N_q$ scaling per order (\figurename~\ref{fig:generic-order}) which quickly becomes intractable.
The inversion of the $\calL$ operator in $\bm q$-space is the best approach compared to any perturbative expansion of diagrams.

\begin{table*}[hbtp]
\centering
    \setlength{\tabcolsep}{1.95cm}
\begin{tabularx}{\linewidth}{@{\hspace{0pt}}lcc}
\hline\hline
Code & Time & Memory \\
\hline
Legacy SSCHA ($\dfour$) & $\mathcal{O}(N_\text{uc}^6 \bm{N_q^6} + n_cN_\text{uc}^4N_q^4)$ & $\mathcal{O}(N_\text{uc}^4 {N_q^4})$ \\
Legacy SSCHA (Bubble approx.) & $\mathcal{O}(N_\text{uc}^3\bm{N_{q}^4} + n_cN_\text{uc}^3N_q^3)$ & $\mathcal{O}(N_\text{uc}^3 {N_q^3})$ \\
Best theoretical Bubble approx. & $\mathcal{O}(N_\text{uc}^3\bm{N_{q}^2})$ & $\mathcal{O}(N_\text{uc}^3 {N_q^2})$ \\
TD-SCHA Lanczos\cite{monacelli_time-dependent_2021} & $\mathcal{O}(n_c N_\text{uc}^3 \bm{N_q^4})$ & $\mathcal{O}(n_c N_\text{uc} N_q)$ \\
\textbf{This work} & $\mathcal{O}(n_c N_\text{uc}^3 \bm{N_q^2})$ & $\mathcal{O}(n_c N_\text{uc} N_q)$\\
\hline\hline
\end{tabularx}
\caption{Computational scaling for the full free energy Hessian. $N_q$: number of $q$-points, $N_\text{uc}$: number of atoms in the primitive cell, $n_c$: stochastic configurations. All iterative methods assume a system-size-independent number of solver steps. Notably, both the TD-SCHA Lanczos and this work are the only references where the cost of computing the bubble or the full RPA has exactly the same scaling, as they are differentiated by considering or neglecting $w[\bm{\Upsilon^{(1)}_q}]$ inside \eqname~\eqref{eq:Lapply:q:Y}, which shares the same scaling with the bubble only calculation. For comparison with other algorithms by different codes, see \appendixname~\ref{app:scaling:codes}.}
\label{tab:scaling}
\end{table*}

\begin{figure}[tbp]
    \centering
    \begin{tikzpicture}[
        phonon/.style={decorate, decoration={snake, amplitude=0.7mm,
            segment length=2.8mm, pre length=0.8mm, post length=0.8mm}, thick},
    ]
    \coordinate (LA) at (-1.9, 0.00);   
    \coordinate (LT) at (-1.1, 0.55);   
    \coordinate (LB) at (-1.1,-0.55);   
    \coordinate (RA) at ( 1.9, 0.00);   
    \coordinate (RT) at ( 1.1, 0.55);   
    \coordinate (RB) at ( 1.1,-0.55);   

    \draw[phonon] (-3.4, 0) -- (LA)
        node[midway, above=1.5mm] {$\bm{q}$};
    \draw[phonon] (RA) -- ( 3.4, 0)
        node[midway, above=1.5mm] {$\bm{q}$};

    \draw[phonon] (LT) to[out=55,  in=125]
        node[midway, above=1.5mm] {$\bm{q}_1$} (RT);
    \draw[phonon] (LB) to[out=-55, in=-125]
        node[midway, below=1.5mm] {$\bm{q}_2$} (RB);

    \filldraw[fill=white, draw=black, thick] (LA) -- (LT) -- (LB) -- cycle;
    \filldraw[fill=white, draw=black, thick] (RA) -- (RT) -- (RB) -- cycle;

    \node at (-1.37, 0) {\small$\bm{\dthree}$};
    \node at ( 1.37, 0) {\small$\bm{\dthree}$};

    \node at (0, 0) {$\bm{\Lambda}$};

    \node at (0, -2.0) {$\bm{q}_1 + \bm{q}_2 = \bm{q} + \bm{G}$};
    \end{tikzpicture}

    \caption{Diagrammatic representation of the SSCHA phonon's bubble term $\bm \dthree \bm \Lambda \bm \dthree$. An incoming phonon of quasi-momentum $\bm{q}$
    decays into two intermediate phonons $\bm{q}_1$ and $\bm{q}_2$, which
    recombine at the second vertex into an outgoing phonon of the same
    quasi-momentum, with $\bm{q}_1 + \bm{q}_2 = \bm{q} + \bm{G}$.}
    \label{fig:bubble}
\end{figure}
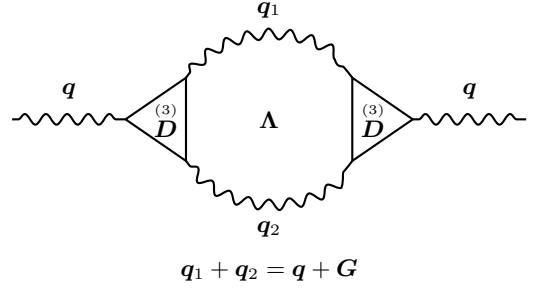

\section{Benchmarking and scaling}

\subsection{Scaling analysis}

\label{sec:benchmark}
\begin{figure*}[tbp]
    \centering
    \includegraphics[width=\textwidth]{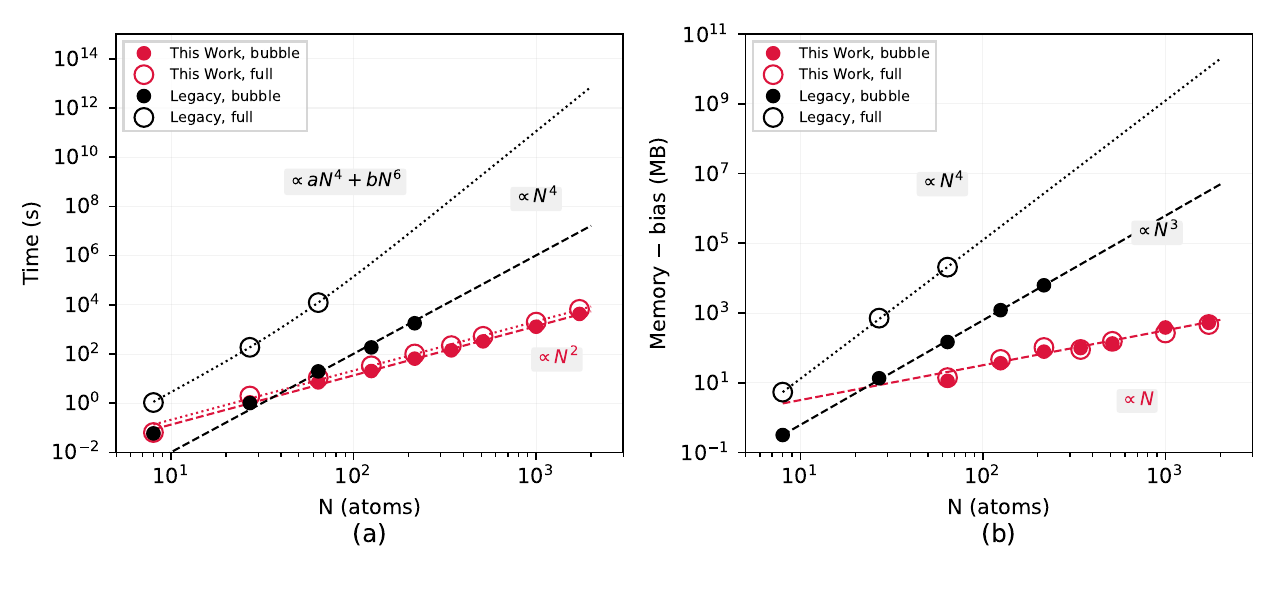} 
    \caption{Computational time and memory peak consumption required to evaluate the free-energy Hessian as functions of the number of atoms in a FCC gold crystal (one atom per primitive cell), for both the bubble approximation (\eqname~\ref{eq:free:energy:hessian:bubble}) and the full RPA inversion (\eqname~\ref{eq:free:energy:hessian}). Black markers and lines correspond to the direct (legacy) formulation, which is extremely prohibitive to use for large systems since it requires the explicit construction of third- and fourth-order tensors. Red markers and lines correspond to the q-space formulation introduced in \secname~\ref{sec:qspace}, demonstrating the qualitative change in scaling laws achieved by the novel approach without introducing any additional approximation (the results are comparable to machine precision for the system size reachable for the legacy algorithm). The observed scaling laws are consistent with those reported in Table~\ref{tab:scaling}. All data points were obtained for FCC gold using ensembles of $n_c=400$ configurations at $T=100$K. Peak memory usage is estimated by monitoring the process during the evaluation of the Free Energy Hessian (details in \appendixname~\ref{Appendix:Scaling:Analysis}). The comparison to algorithms developed in similar legacy codes is discussed in \appendixname~\ref{app:scaling:codes}.}
    \label{fig:scaling}
\end{figure*}

In Figure~\ref{fig:scaling}, we report the measured computational time and memory scaling with system size for the free-energy-Hessian calculation in an FCC gold structure. We compare the direct evaluation with the new linear-system method in momentum space presented in this work (section \ref{sec:linear:system}). For each method, we show two curves: one corresponding to the bubble approximation (\eqname~\ref{eq:free:energy:hessian:bubble}), and one to the full fourth-order expression (\eqname~\ref{eq:free:energy:hessian}).

Since FCC Gold has one atom per primitive cell, the number of atoms coincides with the number of $q$-points: $N=N_q$. The measured time and memory scaling is consistent with the theoretical estimates in Table \ref{tab:scaling}, and demonstrates the fundamental importance of the Q-space formulation for simulating large systems that are entirely inaccessible through the direct approach. 

The direct evaluation based on Eqs. \eqref{eq:d3:new} and \eqref{eq:d4:new}, implemented in the SSCHA code until now (legacy)\cite{bianco_second-order_2017, monacelli_stochastic_2021}, requires the full materialization of $\bm \dthree$  and $\bm \dfour$ tensors, which needs respectively $\mathcal O(N_{\bm q}^3)$ and $\mathcal O(N_{\bm q}^4)$ memory. To compute each element of the tensor, we perform a stochastic average over the ensemble of $n_c$ configurations. So the time needed to materialize the tensors scales as $\mathcal O(n_c N_{\bm q}^3)$ and $\mathcal O(n_c N_{\bm q}^4)$. This is also consistent with the calculation of the same quantity by different approaches, like fitting from an ensemble of configurations a polynomial potential, as implemented in TDEP\cite{knoop_tdep_2024}, ALAMODE\cite{tadano_anharmonic_2014}, hiPhive\cite{eriksson_hiphive_2019-1} (more details about comparison with other legacy approaches are in \appendixname~\ref{app:scaling:codes}).

Remember that computing the full free energy Hessian directly involves the inversion of the rank-4 tensor
\begin{equation}
[\mathbbm{1} - \bm{\dfour}\bm\Lambda]^{-1},
\label{eq:inversion}
\end{equation}
where we interpret the inversion by grouping indices two-by-two to transform a rank-4 tensor in a rank-2 matrix, i.e. $\dfour_{(ab)(cd)}$ becomes a matrix of dimensions $(3N_{\bm q})^2\times (3N_{\bm q})^2$.
Inverting \eqname~\eqref{eq:inversion} requires $\mathcal O[(N_{\bm q}^2)^3] = \mathcal O(N_{\bm q}^6)$ operations. For the system sizes considered in the benchmark, the stochastic construction of $\bm{\dfour}$, which also carries the factor $n_c$, may remain a significant or even dominant contribution to the total computational time.  The crossover between the $\mathcal O(n_cN_{\bm q}^4)$ from $\bm\dfour$ materialization and the $\mathcal O(N_{\bm q}^6)$ arising from the dense matrix inversion is investigated in \appendixname~\ref{Appendix:Scaling:Analysis}.

Conversely, without any further approximation, the q-space approach introduced in \secname~\ref{sec:qspace} has a time complexity of $\mathcal O(N_{\bm q}^2)$ and linear memory $\mathcal O(N_{\bm q})$. Notably, the same scaling applies to both the bubble approximation and the full Hessian, as they differ only by a small prefactor (\figurename~\ref{fig:scaling}). 

The impact of this reduction in terms of memory and time costs is dramatic. Extrapolating the legacy algorithm to the same FCC gold crystal with 1728 atoms (a $12\times12\times12$ supercell) would require approximately $5$ PB of RAM and more than $300.000$ years of computational time. A calculation started by the first Neanderthals would still be running today. In contrast, using the new $q$-space linear-system formulation, we evaluated the same quantity on a laptop using only $1$ GB of RAM and approximately 3 hours of computational time.

\subsection{Displacive phase transition in metal-halide perovskite \ch{CsSnI3}}
\label{sec:cssni3}
We now demonstrate the capabilities of the approach on a realistic, strongly anharmonic system where converged free-energy Hessians were previously out of reach.

\ch{CsSnI3} is a lead-free metal-halide perovskite of great interest for photovoltaic and optoelectronic applications. Above $\sim\SI{426}{\kelvin}$ it adopts the cubic $Pm\bar3m$ black-perovskite structure (B-$\alpha$ phase), which on cooling transforms through a sequence of displacive transitions driven by octahedral tilting from cubic (B-$\alpha$), tetragonal (B-$\beta$) to finally orthorhombic (B-$\gamma$)\cite{Chung2012,monacelli_first-principles_2023}. These transitions are governed by zone-boundary soft modes at the $M=(\tfrac12\tfrac12 0)$ and $R=(\tfrac12\tfrac12\tfrac12)$ points of the cubic Brillouin zone, whose energy is renormalized by anomalously large thermal ionic fluctuations\cite{monacelli_first-principles_2023,monacelli_analyzing_2025}. Resolving the onset of these instabilities requires evaluating the free-energy Hessian on a $q$-mesh dense enough to sample the zone boundary and its surroundings. Crucially, the direct inversion of \eqname~\eqref{eq:free:energy:hessian} is in practice limited to the $2\times2\times2$ supercell (\num{40} atoms), which commensurates only the bare $M$ and $R$ points and cannot resolve the dispersion of the unstable branches; a severe finite-size limitation for a critical phenomenon whose correlation length grows near the transition.

We model the Born--Oppenheimer (BO) surface with a NequIP machine-learning interatomic potential (MLIP) trained on PBEsol DFT data. The details of the training set, model parameters, and quality of the fit are discussed in \appendixname~\ref{app:fit}.

\begin{figure*}[htbp]
    \centering
    \includegraphics[width=\textwidth]{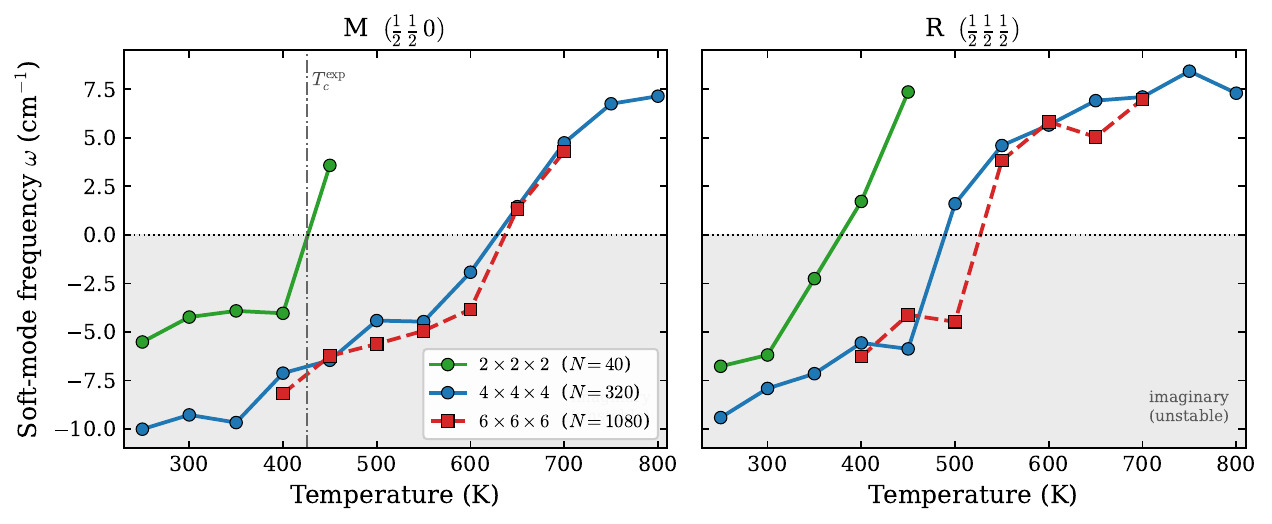}
    \caption{Convergence of the \ch{CsSnI3} free-energy Hessian with supercell size. Lowest square-root eigenvalue of the full (four-phonon) free-energy Hessian at the $M$ (left) and $R$ (right) zone-boundary points of the cubic $Pm\bar3m$ phase, as a function of temperature, for the $2\times2\times2$ ($N=\num{40}$), $4\times4\times4$ ($N=\num{320}$), and $6\times6\times6$ ($N=\num{1080}$) supercells. Negative values (shaded region) denote imaginary frequencies, i.e.\ a structural instability of the cubic phase (negative Hessian eigenvalue). The small $2\times2\times2$ cell---the practical limit of the direct inversion of \eqname~\eqref{eq:free:energy:hessian}---strongly overestimates the phase stability, whereas the $4\times4\times4$ and $6\times6\times6$ results converge onto each other. The dash-dotted line marks the experimental cubic-to-tetragonal transition temperature\cite{monacelli_first-principles_2023}.}
    \label{fig:softmodes}
\end{figure*}

Exploiting the new $q$-space algorithm, we compute the \emph{full} free-energy Hessian of cubic \ch{CsSnI3} retaining the four-phonon scattering (i.e., overcoming the \emph{bubble} approximation) as a function of temperature on supercells up to $6\times6\times6$ (\num{1080} atoms, \num{3240} modes, $N_q=216$), using a stochastic ensemble of $\sim\num{32000}$ configurations. \figname~\ref{fig:softmodes} reports the lowest Hessian frequency (square root of the eigenvalue) at $M$ and $R$ versus temperature for the $2\times2\times2$ ($N=\num{40}$), $4\times4\times4$ ($N=\num{320}$), and $6\times6\times6$ ($N=\num{1080}$) supercells.

The supercell dependence is dramatic: the $2\times2\times2$ cell (i.e., the only size accessible to direct inversion) strongly overestimates the stability of the cubic phase, predicting both soft modes to harden through zero already near $\sim\SI{450}{\kelvin}$. Enlarging the cell to $4\times4\times4$ softens the modes by up to $\SI{5}{\per\centi\metre}$, bringing the transition temperature to about \SI{600}{\kelvin}, while the $6\times6\times6$ data confirm that the $4\times4\times4$ result is essentially converged, the residual difference being comparable to the stochastic noise. This level of convergence of the four-phonon free-energy Hessian on a kilo-atom supercell, impossible with any previously available method, is the central result of this work. 

Unfortunately, achieving proper supercell convergence reveals that the agreement with experiments reported in previous work\cite{monacelli_first-principles_2023} likely emerges from a fortuitous error cancellation between the GGA exchange-correlation potential and finite-size effects. In fact, semi-local functionals are well-documented to overestimate the energy barrier for octahedral tilting in metal halide perovskites\cite{Kaiser2021}, thereby overestimating the transition temperatures compared to experiments. Consequently, these results underscore the need to revise the \ch{CsSnI3} phase diagram by combining accurate hybrid functionals with the large-scale supercells enabled by this work.

\begin{figure*}[htbp]
    \centering
    \includegraphics[width=\textwidth]{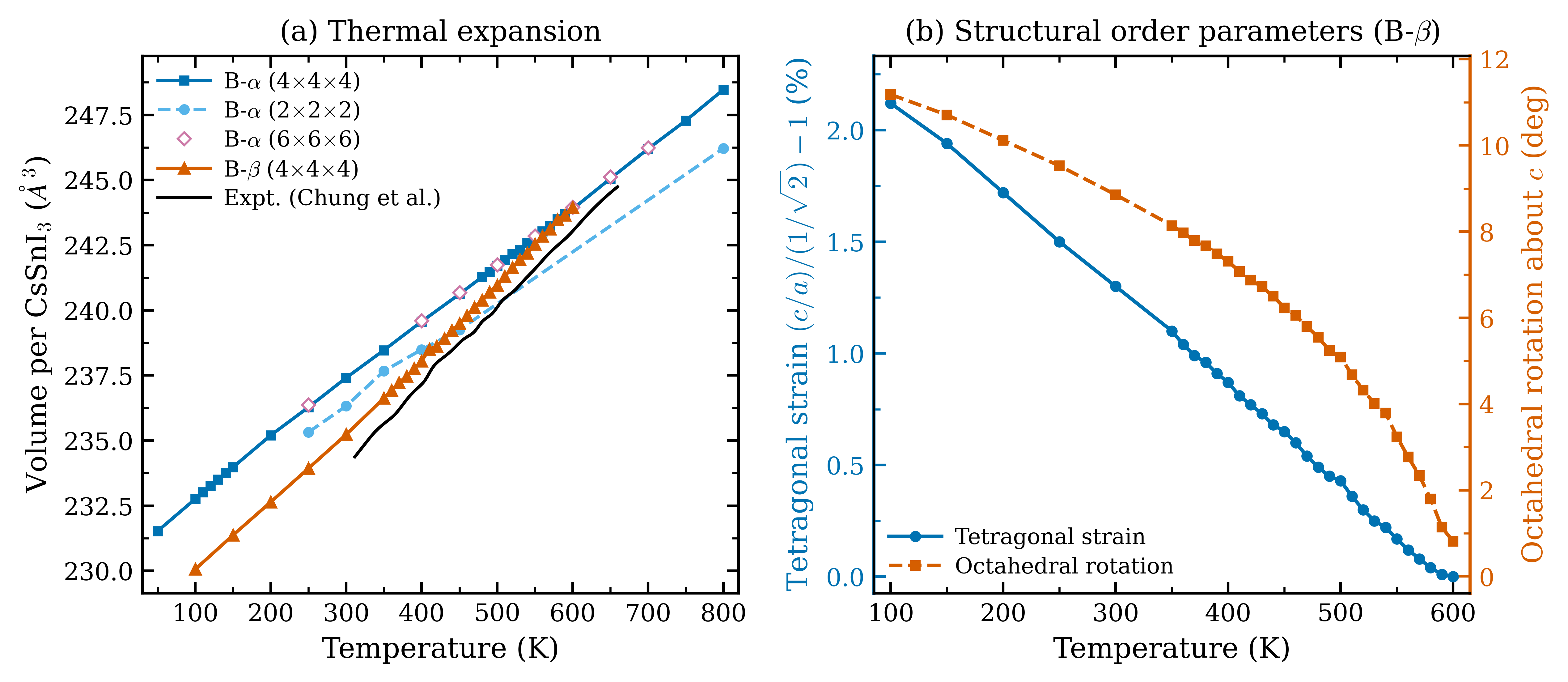}
    \caption{Structural changes of the cubic (B-$\alpha$) and tetragonal (B-$\beta$) phase of perovskite \ch{CsSnI3}. \textbf{(a) Thermal expansion}. The cubic lattice thermal expansion shows that it is well converged with a 4x4x4 supercell (corresponding to 320 atoms) and completely overlaps with the 6x6x6 supercell (1080 atoms). The tetragonal B-$\beta$ volume merges with the cubic one at \SI{600}{\kelvin}. The data are compared with experimental data by Chung et al\cite{Chung2012} (figure S4). \textbf{(b) Structural parameters of B-$\beta$ phase}. Both the strain deformation from cubic and the tilting angle of the octahedra converge toward the expected cubic value (0) at \SI{600}{\kelvin}, confirming that it is the converged phase-transition temperature. The B-$\beta$ 4x4x4 supercell contains 640 atoms, as its primitive cell is twice the cubic one. 
    }
    \label{fig:volume}
\end{figure*}

While our results agree up to double-precision rounding to the legacy algorithm on the small 40-atom cells, we cannot directly compare the two approaches on the 320- and 1080-atom systems, as they are out of reach for the legacy algorithm. 
Thus, we tracked the phase transition by relaxing the low-symmetry phase (B-$\beta$) within SSCHA, monitoring the order parameter as a function of temperature (lattice vectors and octahedral tilting). \figname~\ref{fig:volume}(a) compares the thermal expansion of the B-$\alpha$ and B-$\beta$ phases as a function of the supercell size: above the 4x4x4 q-mesh, the volume per formula unit is essentially independent of supercell size. The B-$\beta$ volume expansion converges toward the B-$\alpha$ at about \SI{600}{\kelvin}, in good agreement with the temperature at which the B-$\alpha$ phase becomes stable in the $M$ point. \figurename~\ref{fig:volume}(b) reports the two structural parameters related to this phase transition: the tetragonal cell strain ($c/a$) and the tilting of the $I$ octahedra around the $c$ axis. Both parameters converge clearly to the cubic value at about \SI{600}{\kelvin}, providing an additional independent quality check of the phase transition estimated from the free energy Hessian.

\begin{figure}
    \centering
    \includegraphics[width=\linewidth]{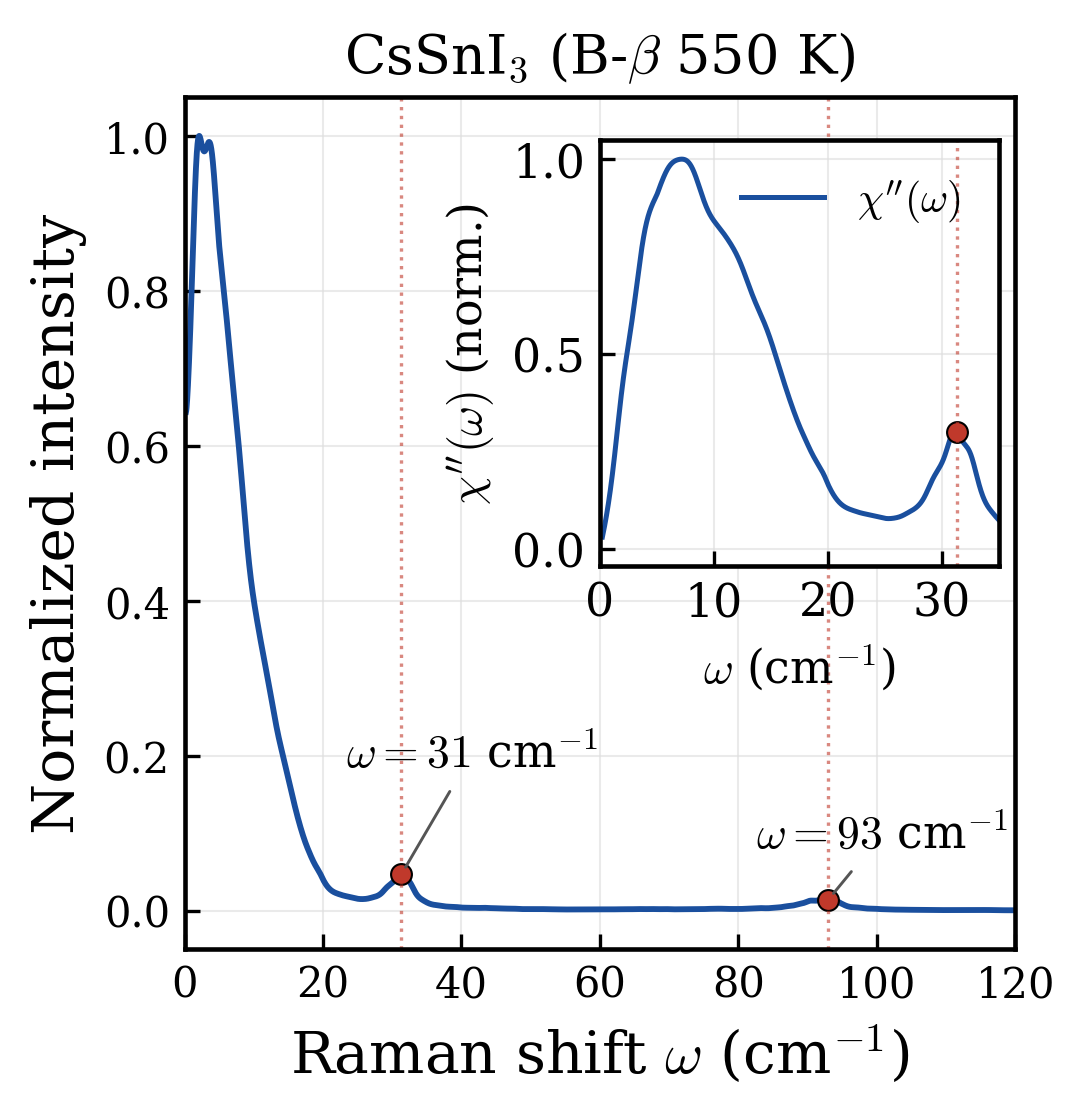}
    \caption{Simulated power-average Raman spectrum of the B-$\beta$ phase of \ch{CsSnI3} at \SI{550}{\kelvin} (Stokes). The $\beta\to\alpha$ transition occurs at $\sim$\SI{600}{\kelvin} within the DFT exchange-correction employed. The soft phonon mode is responsible for most of the signal visible in Raman. In the inset, we report the imaginary part of the polarizability correlation function, which does not account for the Bose scattering peak of the Stokes, enhancing the low-frequency region of the spectrum. The soft mode appears as a large, featureless boson-like peak, evidencing the overdamped nature of the oscillation and its strongly non-Lorentzian shape due to the anharmonicity near the critical point.}
    \label{fig:tdscha}
\end{figure}

The method introduced here can also be applied to dynamical response functions. Indeed, the efficient evaluation of the $\calL^{\text{(anh)}}$ operator represents the main bottleneck of the Lanczos algorithm within the time-dependent stochastic self-consistent harmonic approximation (TD-SSCHA)\cite{siciliano_wigner_2023}. To demonstrate the effectiveness of this approach, we compute the Raman spectrum of \ch{CsSnI3} just below the $\beta\to\alpha$ transition (at \SI{550}{\kelvin}), where the soft mode responsible for the symmetry breaking at $M$ in the high-temperature cell becomes Raman active in the broken symmetry phase B-$\beta$ (\figurename~\ref{fig:tdscha}). 
This soft mode is strongly anharmonic and accounts for the broad peak observed in the unpolarized Raman spectrum. In the Stokes scattering regime, the Raman spectrum is calculated from the polarizability correlation function $C(\omega)$, which is related to the imaginary part of the polarizability response function via the fluctuation-dissipation theorem:
\begin{equation}
    C(\omega) = -\frac{1 + n(\omega)}{\pi}\Im \chi_\text{unpol}(\omega).
\end{equation}
The unpolarized Raman response function $\chi_{\text{unpol}}(\omega)$ is obtained by orientational averaging over the isotropic and anisotropic invariants, as discussed in \appendixname~\ref{app:raman}.

\section{Discussion}

There are multiple approaches to compute the static and dynamical response functions of highly anharmonic materials. SSCHA is an approximate framework that constrains the ansatz for the (dynamical) density matrix to a Gaussian. 
The ground truth can be computed, in principle, with path-integral molecular dynamics in the static limit, while for exact quantum dynamical response functions things are much more challenging, as exact quantum methods are hampered by the phase problem, which causes the estimation to scale exponentially. These approaches are extremely computationally heavy and challenging when large-scale simulations are required to converge finite-size effects, such as in the presence of displacive phase transitions.
To solve this problem, multiple alternatives have been developed. The temperature-dependent effective potential (TDEP) method can fit the static response functions to classical or quantum MD trajectories\cite{knoop_tdep_2024,hellman_lattice_2011,castellano_mode-coupling_2023,thomas_predicting_2010,fransson_dynasortool_2021}, while self-consistent phonons (SCP)\cite{tadano_anharmonic_2014,tadano_self-consistent_2015,eriksson_hiphive_2019-1,lin_first-principles_2026} is a mean-field approach similar to the SSCHA, where a phonon self-energy is solved self-consistently. 
While a detailed comparison across methods is beyond the scope of this work, as already deeply discussed in ref.~\cite{monacelli_analyzing_2025}, we just briefly summarize what differentiates the SSCHA: i) SSCHA is a quantum theory, thus correctly capturing quantum nuclear fluctuations, crucial in the presence of light atoms, strong covalent bonds, or at cryogenic temperatures (in contrast to methods that rely on classical MD). ii) The SSCHA does not approximate the BO energy landscape, which retains its full Taylor expansion. Therefore, the SSCHA automatically includes high-order terms, as the phonon-scattering vertices are dressed with higher-order anharmonicity, and always accounts for mixing of polarization vectors (mode-mixing) and anharmonic-driven structural relaxation (the centroids $\bm {\mathcal R}$ and the lattice via Gibbs free energy minimization\cite{monacelli_pressure_2018}). As a consequence, high-order diagrams arising from dressing vertices, often neglected in SCP (like the quintic-third-order bubble\cite{xia_quintic-anharmonicity-assisted_2026}) or dynamical TDEP~\cite{castellano_mode-coupling_2023} (like the full dynamical RPA self-energy from \figurename~\ref{fig:generic-order}) are automatically included.

The resulting frequency-dependent self-energy allows the SSCHA to describe system spectral functions that deviate from a Lorentzian shape, even when the phonon quasiparticle picture breaks down. 
In contrast, SSCHA (and its dynamical extension\cite{monacelli_time-dependent_2021,lihm_gaussian_2021}) is limited to Gaussian fluctuations, i.e., the centroids and cross-phonon covariance matrix, which correspond to one-phonon and two-phonon propagators with mode-mixing. Therefore, dynamical diagrams where we have more than two phonon lines propagating at the same time (vertical cut) are forbidden. Among them, the Saturn diagram~\cite{siciliano_wigner_2023} is not included in the SSCHA self-energy.

We have removed the principal computational barrier to evaluating the full SSCHA free-energy Hessian in periodic crystals. The novel linear-system momentum-space formulation scales linearly with the $\bm q$-mesh for each perturbation, enabling the simulation of large-scale systems and systematically overcoming the bubble approximation. This is not only extremely beneficial for identifying second-order phase transitions, which is the main focus of this work, but also enables the calculation of dynamical response functions within the TD-SCHA, as the same reformulation also applies as it is to the time-dependent Lanczos algorithm.
Therefore, this work paves the way for simulating phase diagrams and spectral properties in strongly anharmonic materials, where quantum fluctuations play a major role, and supercell-based methods like Path-Integral Molecular Dynamics are computationally prohibitive for dense $\bm q$-point grids. A prime example is high-pressure hydrogen, where the anharmonicity is extremely strong, and the bubble approximation has been demonstrated to fail\cite{monacelli_black_2020}; hydrogen-bond symmetrization in ice\cite{cherubini_quantum_2024} and clathrates\cite{monacelli_hydrogen_2025}. It opens a direct route to phase boundaries and momentum-resolved vibrational spectra in systems where four-phonon physics and long correlation lengths must be treated together; the computational wall that once defined the accessible physics has become a controllable convergence parameter.

\section*{Acknowledgments}
The authors acknowledge EUROHPC and CINECA under the ISCRA initiative for access to HPC resources.

\section*{Code availability statement}

The algorithm described in this work is implemented in the open-source \texttt{tdscha} package\cite{monacelli_stochastic_2021,monacelli_time-dependent_2021}, and will be published under the GPLv3 license in version \texttt{1.7} of the SSCHA software suite (planned for autumn 2026). Starting from version \texttt{1.7}, this will be the default approach in the \texttt{python-sscha} package for free-energy Hessian calculations, with an opt-out strategy allowing the user to revert to the legacy algorithm. Documentation, tutorials, and news about future schools are regularly updated on the SSCHA website (\href{www.sscha.eu}{www.sscha.eu}).

\section*{Data availability statement}
All the data represented in this work are available from the corresponding author upon reasonable request.

\appendix

\section{Mass-rescaled quantities}
\label{app:mass:rescale}

To simplify equations, it is convenient to work in mass-rescaled quantities, as introduced in ref.~\cite{monacelli_time-dependent_2021}.
In particular, all stems from the definition of mass-rescaled position, as done in \eqname~\eqref{eq:mass:rescale}.
This converts to forces in the following way:
\begin{equation}
    \mathrm f_a = -\frac{dV}{dR_a} = -\frac{1}{\sqrt{m_a}}\frac{dV}{dr_a} = \frac{{\mathbbm f}_a}{\sqrt{m_a}}
\end{equation}
where $\mathbbm f_a$ is the standard Born-Oppenheimer force, i.e., the derivative of the ionic energy landscape with respect to the atomic position. Analogously, the force constants matrices are defined as
\begin{equation}
    \Phi_{ab} = \frac{d^2V}{dr_adr_b},\qquad
    \Dthree_{abc} = \frac{d^3 V}{dr_adr_bdr_c},
\end{equation}
\begin{equation}
    \Dfour_{abcd} = \frac{d^4V}{dr_adr_bdr_cdr_d}.
\end{equation}
The relationship to the $\bm D$, $\bm\dthree$, and $\bm\dfour$ defined in the paper is
\begin{equation}
    D_{ab} = \frac{\Phi_{ab}}{\sqrt{m_am_b}},\qquad
    \dthree_{abc} = \frac{\Dthree_{abc}}{\sqrt{m_am_bm_c}},
\end{equation}
\begin{equation}
    \dfour_{abcd} = \frac{\Dfour_{abcd}}{\sqrt{m_am_bm_cm_d}}.
\end{equation}
In the same way, the mass-rescaled inverse covariance matrix $\bm{\Upsilon}$ is defined in the following way:
\begin{equation}
    \varPsi_{ab} = \avg{(r_a- r^{(0)}_a)(r_b-r^{(0)}_b)}\;\;
    \Psi_{ab} = \avg{u_au_b} = \frac{\varPsi_{ab}}{\sqrt{m_am_b}}
\end{equation}
then, the mass-rescaled inverse covariance matrix is
\begin{equation}
\bm{\Upsilon} = \left(\bm\Psi\right)^{-1} .
\end{equation}

\section{Equivalence proof}\label{app:equivalence:proof}

Here we prove that the solution of the linear system \eqname~\eqref{eq:linear:system} yields the free energy Hessian as in \eqname~\eqref{eq:static:G}. The explicit linear system in \eqname~\eqref{eq:linear:system} that we need to solve for $\bm{X^k}$ and $\bm {Y^k}$ is
\begin{equation}
    \begin{cases}
            \displaystyle \bm{D} \bm{X^k} + \bm{\dthree}\bm {Y^k} = \bm{\delta_k} \\
            \\
            \displaystyle
        \bm{\dthree}\bm{X^k} + {\bm \dfour} {\bm Y^{k}} - \bm {\Lambda}^{-1}\bm {Y^k}  = 0 
    \end{cases}.
    \label{eq:block:system}
\end{equation}

We first find $\bm Y^{k}$ from the second equation of~\eqref{eq:block:system}
$$
\left({\bm \dfour}  - \bm\Lambda^{-1}\right)\bm {Y^k} = -\bm{\dthree} \bm{X^k}
$$
\begin{equation}
    \bm {Y^k} = -\left({\bm \dfour}  - \bm\Lambda^{-1}\right)^{-1}\,\bm{\dthree}\bm{X^k}.
    \label{eq:upsilon:solved}
\end{equation}
Substituting in the first equation, we get
\begin{equation}
    \left[\bm{D} -\bm{\dthree}\left({\bm \dfour}  - \bm\Lambda^{-1}\right)^{-1} \bm{\dthree}\right]\bm{X^k} = \bm{\delta_k},
    \label{eq:schur:complement}
\end{equation}
which can be solved by inverting the term in the parentheses:
\begin{equation}
    \bm{X^k} = \left[\bm{D} -\bm{\dthree}\left({\bm \dfour}  - \bm\Lambda^{-1}\right)^{-1} \bm{\dthree}\right]^{-1}\bm{\delta_k}.
\end{equation}
Since the susceptibility matrix $\bm \calG$ has $\bm {X^k}$ as $k$-th column, the full matrix is obtained replacing the $\bm{\delta_k}$ vector with an identity matrix, therefore
\begin{equation}
    \bm\calG = \left[\bm{D} -\bm{\dthree}\left({\bm \dfour}  - \bm\Lambda^{-1}\right)^{-1} \bm{\dthree}\right]^{-1}.
\end{equation}
Since from \eqname~\eqref{eq:hessian:inverse} the positional free energy Hessian is the inverse of $\bm\calG$, then we have
\begin{equation}
    {\bm \calG}^{-1} = \bm{D} -\bm{\dthree}\left({\bm \dfour}  - \bm\Lambda^{-1}\right)^{-1} \bm{\dthree}\label{eq:F:almost}
\end{equation}
which is very similar to the expression of \eqname~\eqref{eq:free:energy:hessian} derived in \cite{bianco_second-order_2017}. In fact, they are exactly equivalent, as
\begin{align}
    \left(\bm {\dfour} - \bm\Lambda\right)^{-1} &=  -\left(\bm{\Lambda}^{-1} - \bm{\dfour}\right)^{-1} \nonumber\\
    &= -\left[(\mathbbm{1} - \bm{\dfour}\bm{\Lambda})\bm{\Lambda}^{-1}\right]^{-1} \nonumber\\
    &= -\bm{\Lambda}(\mathbbm{1} - \bm{\dfour}\bm{\Lambda})^{-1}.
    \label{eq:L:upsilon:inverse}
\end{align}
Substituting \eqname~\eqref{eq:L:upsilon:inverse} into \eqname~\eqref{eq:F:almost}, we recover \eqname~\eqref{eq:free:energy:hessian}
\begin{align}
    \bm{\calG}^{-1} &= \bm{D} + \bm{\dthree}\bm{\Lambda}(\mathbbm{1} - \bm{\dfour}\bm{\Lambda})^{-1}\bm{\dthree} = \frac{d^2\mathcal F}{d\bm R d\bm R}.
\end{align}

Notably, the variable $\bm Y^{k}$, obtained from the solution of the linear system and discarded, has a physical meaning: it represents the response of the two-phonon covariance matrix to a perturbation of the average atomic positions $k$, capturing how quantum fluctuations relax when the structure is deformed.

\section{Equivalence proof using diagramatic formalism}

Here we present the equivalence proof by using the diagrammatic formulation of SCHA~\cite{siciliano_wigner_2023} introduced in Figs.  \ref{fig:rpa}, \ref{fig:decay}, \ref{fig:decay4}. The main advantage of the diagrammatic representation is that it makes the contraction pattern of the indices in third- and fourth-order tensors more transparent and intuitive. In this section, we represent the fourth-order tensor $\bm{\dfour}$ by a red square, $\dsquare$. Its two vertices on the left indicate that the first pair of tensor indices is combined into a single composite index of dimension $(3N)^2$, while the two vertices on the right represent the second pair of indices, which is combined in the same way. So, it is an $(3N)^2\times (3N)^2$ matrix.

The two-phonon propagator $\bm{\Lambda}$, similarly, is represented by a double line, $\doubleline$, whereas the dynamical matrix $\bm{D}$ is represented by an inverse of a single line, $\dline^{-1}$. The inversion comes from the fact that the dynamical matrix is the inverse of the Phonon static Green function.

We represent the third-order tensor $\bm{\dthree}$ by using yellow triangles. A right-pointing triangle, $\triangler$, indicates that the two indices on the left are combined into a single composite index of dimension $(3N)^2$. So that the dimension of $\triangler$ are $(3N)^2\times 3N$. Conversely, a left-pointing triangle, $\trianglel$, indicates that the two indices on the right are combined into the composite index, so that the dimension is $3N\times (3N)^2$. Notice that one is the transpose of the other $\triangler = \left( \trianglel \right)^T$. More generally, the transposition corresponds to a left-right reflection of the associated diagram (exchanging the matrix indices corresponds to flip the two matrix indices, and so flipping the diagram).

For example, take the product $\bm \dthree \bm \Lambda \bm \dthree$. The result is a $3N\times 3N$ matrix, and any element can be written expicitely as:
\begin{equation}
  \left( \bm{\dthree}\bm{\Lambda}\bm{\dthree} \right)_{ab}
  = \sum_{(hk)(lm)} \underbrace{\dthree_{a(hk)}}_{{}_a\trianglel_{h}^k}\, \underbrace{\Lambda_{(hk)(lm)}}_{
   {}^{\raisebox{0.4ex}{$\scriptstyle k$}}_
    {\raisebox{-0.4ex}{$\scriptstyle h$}}
  \mkern3mu
  \vphantom{\doubleline}
  \doubleline
  \vphantom{\doubleline}
  \mkern3mu
  {}^{\raisebox{0.4ex}{$\scriptstyle m$}}_
    {\raisebox{-0.4ex}{$\scriptstyle l$}}
  }\, \underbrace{\dthree_{(lm)b}}_{{}\prescript{m}{l}\triangler_b} ,
\end{equation}
We putted inside the brackets $(\cdot)$ the couples of indices which are combined and flattened (in this case, $(hk)$ and $(lm)$).
This product can be easily written diagramatically as:
\begin{equation}
    \bm \dthree \bm \Lambda \bm \dthree =  \trianglel \doubleline \triangler
\end{equation}
Now it is clear which indices are summed just by looking at the diagrams.

In this formalism, the free energy Hessian in eq. \eqref{eq:free:energy:hessian} is an RPA summation (fig. \ref{fig:rpa}):

\begin{equation}
\begin{aligned}
     \frac{d^2F}{d\bm R d\bm R} &=  
    \dline^{-1} + \trianglel \doubleline \left( \mathbbm 1   - \dsquare \doubleline \right)^{-1} \triangler=\\
    & =\dline^{-1} + \trianglel \doubleline \triangler + 
     \trianglel \doubleline \dsquare \doubleline \triangler + ... 
\label{eqAPP:Free:energy:hessian:Diagram}
\end{aligned}
\end{equation}

The operator $\mathcal L$ reads:
\begin{equation}
    \mathcal L =\begin{pmatrix}
        \dline^{-1} &  \trianglel\\
        \triangler & \dsquare - \doubleline^{-1}
    \end{pmatrix}
\end{equation}
The dimensions of $\mathcal L$ are $3N(1 + 3N) \times 3N(1 + 3N) $.
We saw that the solutions $\ket{\phi_{k}} $of the $3N$ linear systems \[\mathcal L\ket{\phi_{k}} =\begin{pmatrix}\bm \delta_{k}\\\bm 0\end{pmatrix} \text{ , with } k =1,...,3N\] can be separated in two parts $\bm X^{ k}$ which is a vector containing $3N$ elements, and $\bm Y^{k}$, which contains $(3N)^2$ elements. This block structure has a transparent diagrammatic counterpart: 
$\bm X^k$
 is drawn as an object with a single leg, and 
$\bm Y^k$
 as an object with two legs, the number of legs matching the number of phonon indices carried by each block.
\begin{equation}
    \bm X^{k}\equiv \Xkvertex^k \in \mathbb R^{3N}
\end{equation}
\begin{equation}
    \bm Y^{k }\equiv \Ykvertex^k \in \mathbb R^{(3N)^2}
\end{equation}

\begin{equation}
    \begin{pmatrix}
        \dline^{-1} &  \trianglel\\
        \triangler & \dsquare - \doubleline^{-1}
    \end{pmatrix}
\begin{pmatrix}
    \Xkvertex^k\\
    \Ykvertex^k
\end{pmatrix}
= 
\begin{pmatrix}
    \bm \delta_k\\
    \bm 0
\end{pmatrix}
\end{equation}

So, the linear system becomes:
\begin{equation}
    \begin{cases}
        \dline^{-1} \Xkvertex^k + \trianglel \Ykvertex^k = \delta_{k}\\
        \triangler \Xkvertex^k + \left (\dsquare - \doubleline^{-1}\right ) \Ykvertex^k = 0
    \end{cases}
\end{equation}
From the second equation, putting $\mathbbm 1 = \doubleline \doubleline^{-1}$:
\begin{equation}
\begin{aligned}
    \Ykvertex^k& = -\left(\dsquare - \doubleline^{-1}\right)^{-1} \triangler \Xkvertex^k=\\
    & =-    \left[\left(\dsquare \doubleline- \doubleline^{-1}\doubleline\right)\doubleline^{-1}\right]^{-1} \triangler \Xkvertex^k
    =\\
    &=\doubleline
    \left(\mathbbm 1-  \dsquare\doubleline 
    \right)^{-1}\triangler \Xkvertex^k
\end{aligned}
\end{equation}
Putting it inside the first equation:
\begin{equation}
    \underbrace{\left(
    \dline^{-1} + \trianglel \doubleline\left( 1 -\dsquare\doubleline \right)^{-1}
     \triangler
    \right)}_{\frac{d^2F}{d\bm R d\bm R}}\Xkvertex^k
    = \delta_{k}
\end{equation}

So, the different solutions $\Xkvertex^k$ of the $k-$th linear system are the columns of the inverse of the free energy Hessian:
\begin{equation}
    \frac{d^2F}{d\bm R d\bm R} = \begin{pmatrix}
        \Xkvertex^1 &\Xkvertex^2& \cdots &\Xkvertex^{3N}
    \end{pmatrix} ^{-1}
\end{equation}

\section{Iterative solver with harmonic preconditioner}
\label{app:linear:system}
We solve \eqname~\eqref{eq:linear:system} using the GMRES algorithm\cite{monacelli_time-dependent_2021}. The operator $\calL$ is not guaranteed to be positive definite (negative eigenvalues signal a structural instability), so standard conjugate-gradient methods are not applicable in general.

To accelerate convergence, we employ a preconditioner $\bm{M} = (\calL^{(\text{har})})^{-1}$. Since $\calL^{(\text{har})}$ is diagonal in the phonon basis, its inverse is trivial:
\begin{equation}
    \bm{M} = \begin{pmatrix}
        \bm{D}^{-1} & 0 \\
        0 & -\bm{\Lambda}
    \end{pmatrix}
\end{equation}
This preconditioner is exact in the harmonic limit and captures the dominant energy scales, effectively preconditioning away the large frequency spread present near the phase transition, where the frequency of the order parameter shows a softening. Notably, without preconditioning, the condition number $C$ of this system is dominated by $\bm\Lambda$ exactly like the SSCHA minimization\cite{monacelli_pressure_2018}:
$$
C\sim\left(\frac{\omega_\text{max}}{\omega_\text{min}}\right)^\gamma\qquad \begin{array}{lr}
\gamma = 3 & k_bT \ll \omega \\
\gamma = 4 & k_bT \gg \omega
\end{array}.
$$
Therefore, preconditioning is essential to achieve convergence of the GMRES algorithm in few steps. Empirically, we observed in all tests that 10-- 30 steps are required to converge the GMRES algorithm to machine precision, mostly independent of the mesh size, with small variation depending on the nearness of the critical point and whether $\bm{\dfour}$ is accounted for or neglected (if $\bm{\dfour}$ is neglected, we observed systematically less than 10 steps are needed to converge). We do not rule out the possibility that highly anharmonic low-symmetry phases may require more steps to converge, due to the need to relax the polarization vectors compared to the auxiliary SSCHA dynamical matrix.
Notably, this algorithm fully preserves \textbf{mode-mixing}, i.e., polarization vectors are free to change both in the SSCHA minimization and in the calculation of the free energy Hessian.

\section{Fit of the \ch{CsSnI3} Born-Oppenheimer energy landscape}
\label{app:fit}
\begin{figure}[tbp]
    \centering
    \includegraphics[width=\columnwidth]{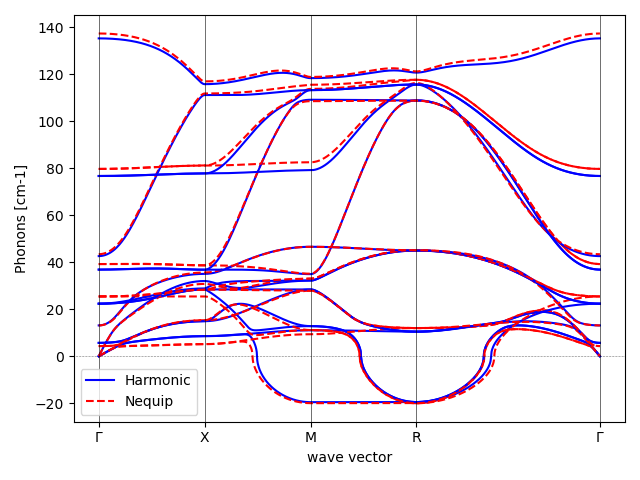}
    \caption{Harmonic phonon dispersion of cubic \ch{CsSnI3} computed with the NequIP MLIP (red dashed) compared to the underlying DFT reference (blue). The potential faithfully reproduces the harmonic spectrum, including the dynamically unstable branches (imaginary frequencies plotted as negative) at the $M$ and $R$ zone-boundary points that drive the displacive transitions.}
    \label{fig:dispersion}
\end{figure}

The BO energy landscape of \ch{CsSnI3} used for the test has been fitted using as training data the first-principles calculations performed in ref.~\cite{monacelli_first-principles_2023}, using the PBEsol exchange-correlation functional with the Quantum ESPRESSO software suite.
The fit is performed with \texttt{nequip} version \texttt{0.6.2}, including in the cost function energies, forces, and stress on the structures sampled from the 40-atom supercell in the B-$\alpha$, B-$\beta$, B-$\gamma$ and Y-$\delta$ phases at various temperatures between \SI{250}{\kelvin} and \SI{450}{\kelvin}. The training set is composed of 1200 independent structures sampled from the SSCHA converged density matrices (computed fully \emph{ab initio} in ref.\cite{monacelli_first-principles_2023}), while the test set consists of 320. The model is an E(3)-equivariant message-passing neural network with four interaction layers, rotational features up to angular order $l_{\max}=1$, 32 independent equivariant features per atom, and a cutoff radius of \SI{5.0}{\angstrom} in which interatomic distances are expanded in a basis of 8 radial functions. The model achieves a force root-mean-square error below \SI{50}{\milli\electronvolt\per\angstrom}. Gated activation functions (SiLU for even, tanh for odd parities) are used as nonlinearities, and the loss function combines forces (unit weight), total energy per atom (unit weight), and stress (weight 0.1). Training uses the Adam optimizer with AMSGrad, a learning rate of \num{0.001} and batch size 8, with early stopping upon validation-loss saturation.
The quality of the fit is independently assessed by investigating the Harmonic phonon, as displayed in \figurename~\ref{fig:dispersion}, where we compare the DFT reference harmonic phonons to the fitted machine-learned interatomic potential (MLIP).

We explicitly verified that, after relaxing with the SSCHA at finite temperatures on the 40-atom cell, the MLIP yields results perfectly consistent with the full DFT reference; both the auxiliary phonon frequencies and free energy Hessian root eigenvalues agree by less than \SI{2}{\per\centi\meter} in the 8 $\bm q$ points commensurate with the 2x2x2 supercell, further validating the fidelity of the machine-learned model for the SSCHA calculation.

\section{Details about the scaling analysis}
\label{Appendix:Scaling:Analysis}

Scaling data which are used to produce Fig. \ref{fig:scaling} come from a measurement of time and memory peak of the process in the call of the only functions responsible of the evaluation of the Hessian in the SSCHA and TDSCHA code. 

All memory data have offsets due to ensemble loaded and preparation. For the legacy algorithm, most of the memory consumption is associated with the Hessian evaluation. In contrast, for the new linear-system-based method, memory usage is mainly determined by Julia’s initialization and dependencies, which require approximately 800 MB of RAM, with fluctuations of about 20 MB caused by the Julia Garbage Collector, and by the ensembles, whose memory scales also linearly with $N$. Consequently, the peak-memory estimate for the proposed algorithm is affected by greater measurement noise. 
Notably, the total measured peak memory usage for the \(12 \times 12 \times 12\) system \((N=1728)\) was only approximately \(1.2\) GB, which is about one order of magnitude lower than the memory required by the legacy algorithm for the much smaller \(4 \times 4 \times 4 \) system \((N=64)\). 

\begin{figure}[tbp]
    \centering
    \includegraphics[width=\linewidth]{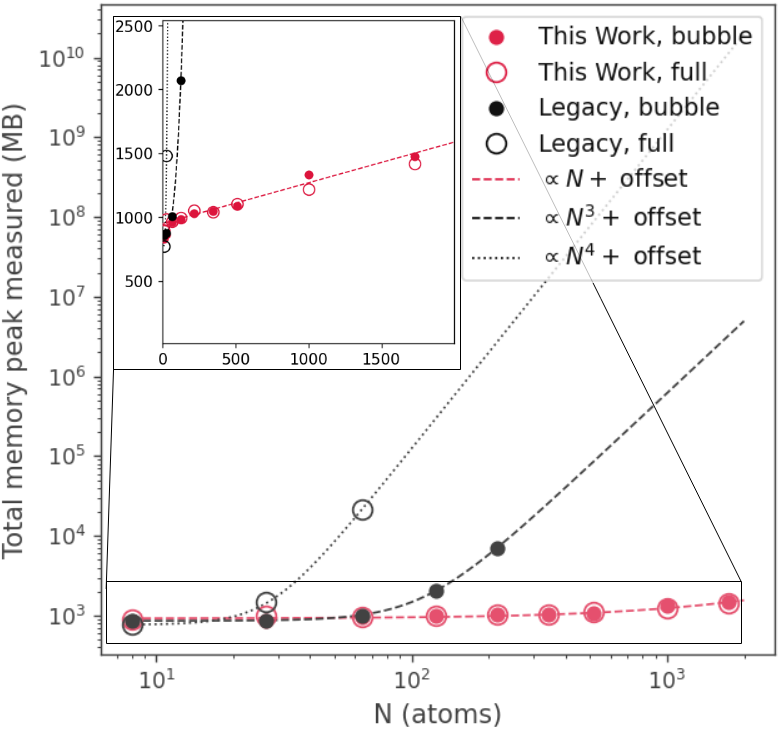}
    \caption{Raw data of the total memory peaks during the evaluation of the Free Energy Hessian with the method presented in this paper. The inset plot is a zoom in linear-linear scale.
    }
    \label{fig:appendix:validation:memory}
\end{figure}

We toggled these offset to see the scaling of the only Hessian evaluation algorithms in the log-log scale in \figname~\ref{fig:scaling}. We present also the raw data of the total memory peak of the process in \figname~\ref{fig:appendix:validation:memory}.

\begin{figure}[tbp]
    \centering
    \includegraphics[width=\columnwidth]{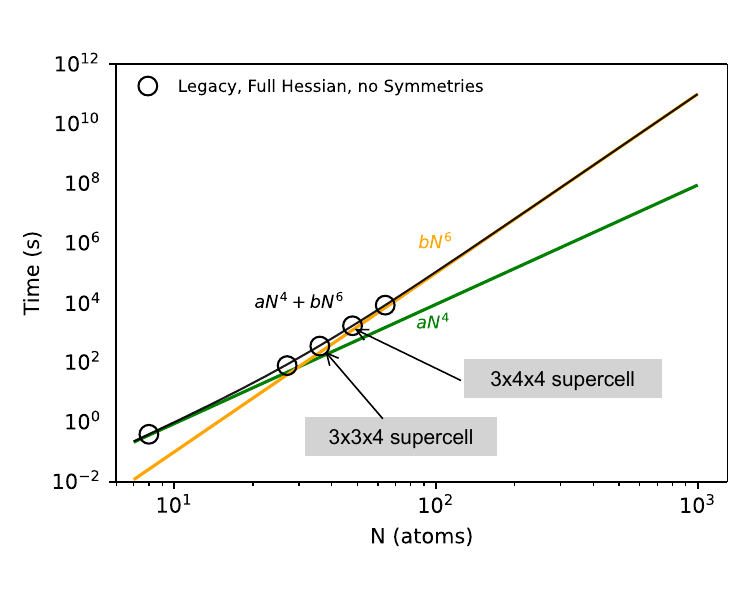}
    \caption{Validation of the $aN^4 + bN^6$ time dependence on 5 data points  of the legacy algorithm in the evaluation of the full Free energy Hessian in a non-cubic Gold. Breaking the cubic symmetries allowed us to add 2 data points between relative to 3x3x4 and 3x4x4 non-cubic supercells. The behavior confirms the correctness of the theoretical prediction in time for the legacy algorithm.}
    \label{fig:appendix:validation:v4}
\end{figure}

The theoretical lines are the result of laws of the type \[y_{a}(N) = aN^\alpha,\]where $\alpha$ is fixed to be the expected scaling. This curve becomes linear in the log-log scale:
\[
\eta_a(\xi) = \log a + \alpha \xi
\]
With $\eta_a \equiv \log y_a$ and $\xi \equiv \log N$. \\
The only exception is in the Legacy full Hessian time evaluated with the legacy algorithm, which is fitted on a non-linear function \[T^{legacy,Full}_{a,b}(N) = aN^4 + b N^6\] Notice that the values of these coefficients depends on the specific system, the number of configurations of the ensembles, and the specific implementation. Since only three data points were computed for this fit, in \figname~\ref{fig:appendix:validation:v4} we provide an additional validation using gold with broken cubic symmetry. This setup allowed us to include two additional data points between the 3$\times$3$\times$3 and the 4$\times$4$\times$4 supercell, providing a more robust validation of the fitted curve and, consequently, of the computational time extrapolation to larger systems.
\\

\section{Scaling of TDEP, ALAMODE and Phono3py codes}
\label{app:scaling:codes}

\begin{table}[hbtp]
    \centering
    \setlength{\tabcolsep}{5mm}
\begin{tabularx}{\linewidth}{@{\hspace{0pt}}lcc}
\hline 
\hline
Code & Time scaling & Memory scaling \\
\hline 
         ALAMODE\cite{tadano_anharmonic_2014} & $\mathcal O(n_c^2N_\text{uc}^5\bm{N_q^4})$ & $\mathcal O(n_cN_\text{uc}^4{N_q^3})$ \\
phono3py\cite{togo_distributions_2015} & $\mathcal{O}(N_\text{uc}^3\bm {N_q^4})$ & $\mathcal O (N_\text{uc}^3N_q^3)$ \\
TDEP$^*$\cite{knoop_tdep_2024}  & $\mathcal O(N_\text{uc}^9\bm {N_q^6})$ & $\mathcal O (N_\text{uc}^6N_q^4)$ \\
\hline
\hline
    \end{tabularx}
    \caption{Scaling with the number of $\bm q$ points of different codes that implement the fitting of the high-order force-constant tensors and the calculation within the bubble approximation (\eqname~\ref{eq:free:energy:hessian:bubble}).}
    \label{tab:code:comparison}
\end{table}
Here, report the scaling with the number of q points $N_q$ of some of the most commonly employed software codes for computing the bubble approximation, and briefly discuss the algorithmic differences with the SSCHA method presented here.

We assume that no approximation is employed on the shape of $\bm\dthree$. In the language of the codes, this means that high-rank force constant matrices are computed without any real-space cutoff, and the parameter reduction only comes from symmetries and acoustic sum rules. Also, we assume that the number of configurations employed, $n_c$, does not grow with $N_q$. This latter assumption is not true in general, as the absence of a cutoff makes the number of unknowns in $\bm\dthree$ grow with the cell size, and therefore, a larger number of force evaluations ($n_c$) is required to determine the $\bm\dthree$ elements with the same accuracy. 

All state-of-the-art codes compute and materialize the high-rank force constant tensors. In particular, for the Bubble approximation, the $\bm\dthree$.
The codes TDEP\cite{knoop_tdep_2024}, ALAMODE\cite{tadano_anharmonic_2014}, and Hiphive\cite{eriksson_hiphive_2019-1} obtain $\bm\dthree$ from fitting the force residuals through a linear system of the form
$$
A\bm x = \bm f
$$
where $\bm x$ are the unknown symmetry-independent entries of $\bm\dthree$, thus it has a length of the order $\mathcal O(N_\text{uc}^3N_q^2)$, while $\bm f$ are the first-principles forces in each configuration inside the ensemble (a vector of size $3N_\text{uc}N_qn_c$).  
The aforementioned codes differ in how the system is solved.
TDEP solves a linear system of the form
$$
A^\dagger A \bm x = A^\dagger \bm f.
$$
This requires materializing $A^\dagger A$, which occupies $\mathcal O(N_\text{uc}^6 N_q^4)$ in memory, and scales as the number of unknowns to the third power in time, i.e., $O(N_\text{uc}^9N_q^6)$.
Its dimensions are from the dimension of all forces in the ensemble times the number of elements in the $\bm\dthree$: $\mathcal O(n_cN_\text{uc}N_q)\times \mathcal O(N_\text{uc}^3 N_q^2)$.

The ALAMODE code\cite{tadano_anharmonic_2014} also builds a fitting table $A$. However, differently from TDEP, it does not form the $A^\dagger A$ matrix, but performs the singular value decomposition (SVD) of the $A$ table directly via LAPACK to invert the system. The computational cost for the SVD decomposition is the square of the row elements times the number of columns: 
$$
\text{SVG: }\qquad 
\mathcal O\left[(n_cN_\text{uc}N_q)^2(N_\text{uc}^3 N_q^2)\right] = 
\mathcal O (n_c^2N_\text{uc}^5N_q^4)
$$
Finally, Hiphive\cite{eriksson_hiphive_2019-1} project the fitting table $A$ on the basis of symmetry allowed constrains (including the ASR). While this approach is very powerful when approximated constraints enforcing sparsity are introduced, it determines an unfavorable $N_q$ scaling when no cutoff is introduced. In fact, the matrix product between $A$ and the constraint map (a rank-2 array of size $(N_\text{uc}^3N_q^2)\times(N_\text{uc}^3N_q^2)$) determines the scaling as $\mathcal O(n_cN_\text{uc}^7N_q^5)$.

All these approaches largely benefit from introducing a sparsity assumption, like a fixed cutoff or a compressed sensing technique, which strongly suppresses the $N_q$ scaling at the cost of a controlled approximation. Notably, such an assumption could also be employed in the SSCHA, further optimizing the current algorithm, but it is beyond the scope of this work.

Then, these codes perform the bubble approximation to compute the free energy Hessian (like ALAMODE\cite{tadano_anharmonic_2014}), or to compute the phonon spectral function (like phono3py\cite{togo_distributions_2015} and TDEP\cite{knoop_tdep_2024}). 
TDEP, ALAMODE and also phono3py compute the bubble diagram in \figurename~\ref{fig:bubble} by iterating over all external $\bm q$ $\mathcal O(N_q)$, all internal $\bm {q_1}$ $\mathcal O(N_q)$, then summing over every real-space atom triplet $\bm\dthree$, which grows, without cutoff, as $\mathcal O(N_\text{uc}^3N_q^2)$ (the specificity on how the $\bm\dthree$ is represented changes in the codes, but the scaling is the same). Therefore, the resulting cost of the Bubble calculation is $O(N_\text{uc}^3 N_q^4)$. Notably, this is the same cost as the SSCHA legacy algorithm in the bubble calculation.
Also in this case, introducing a cutoff could significantly reduce the inner loop, as triplets of atoms far away can be approximated to zero. 

Notably, the TDEP free energy Hessian is exact when extracted from converged path-integral molecular dynamics trajectories, including interactions that go beyond what all other methods here described\cite{monacelli_analyzing_2025}. However, if self-consistent TDEP is used, the $\bm D$ must be corrected with the full Hessian equation in \eqname~\eqref{eq:free:energy:hessian}, like for the SSCHA. Moreover, the bubble approximation is introduced to compute the dynamical spectrum anyway.

It is worth noting that the novel algorithm here introduced not only beats all SOTA codes by at least a factor $N_q^3$ in scaling, but at this cost also goes far beyond the bubble approximation, introducing new physics with a much lower computational cost.

\section{Unpolarized Raman within TD-SSCHA}
\label{app:raman}
The TD-SSCHA code can efficiently compute diagonal response functions, i.e., response functions whose probe and pump observables coincide. For Raman, the response function is the polarizability $\alpha_{xy}$ defined as the derivative of the total energy to an external static electric field:
\begin{equation}
    \alpha_{xy} = \frac{\partial^2\mathcal E}{\partial E_x\partial E_y}
\end{equation}
where $E$ is an external electric field and $\mathcal E$ is the ground-state electronic total energy. The perturbation vector on the TD-SCHA one-phonon channel is
\begin{equation}
    p_k = \sum_{ab}\epsilon^\text{(in)}_a\epsilon^\text{(out)}_b \frac{\partial \alpha_{ab}}{\partial R_k}
\end{equation}
where $\hat {\bm\epsilon}$ is the versor of incoming and outgoing radiation (Raman is a scattering experiment).
The perturbation vector $\bm p$ is used as the initial state $\ket{\psi}$ of the Lanczos algorithm, whose details are reported elsewhere\cite{monacelli_time-dependent_2021}.
The unpolarized Raman response function $\chi_\text{unpol}(\omega)$ can be computed by averaging over the Placzek invariants. We use the following form of Placzek invariants to compute only diagonal perturbations in the Lanczos algorithm: each run employs a symmetry-adapted Raman perturbation vector $\mathbf{d}_k$ ($k=1,\dots,7$), defined in terms of the Cartesian polarization components 
$$
{e_{ij}}_k =  \frac{\partial \alpha_{ij}}{\partial R_k} 
$$
Finally, the 7 perturbations along the Placzek invariants are
\begin{equation}
\mathbf{d}_k = 
\begin{cases}
    \frac{1}{3} (\mathbf{e}_{xx} + \mathbf{e}_{yy} + \mathbf{e}_{zz}), & k = 1, \\[4pt]
    \frac{1}{\sqrt{2}} (\mathbf{e}_{xx} - \mathbf{e}_{yy}), & k = 2, \\[4pt]
    \frac{1}{\sqrt{2}} (\mathbf{e}_{xx} - \mathbf{e}_{zz}), & k = 3, \\[4pt]
    \frac{1}{\sqrt{2}} (\mathbf{e}_{yy} - \mathbf{e}_{zz}), & k = 4, \\[4pt]
    \sqrt{3} \, \mathbf{e}_{xy}, & k = 5, \\[4pt]
    \sqrt{3} \, \mathbf{e}_{xz}, & k = 6, \\[4pt]
    \sqrt{3} \, \mathbf{e}_{yz}, & k = 7.
\end{cases}
\end{equation}
The total unpolarized response function $\chi_{\text{unpol}}(\omega)$ is then expressed as a linear combination of the $7$ diagonal Green's functions $\chi_k(\omega) = \langle \mathbf{d}_k | \calG(\omega) | \mathbf{d}_k \rangle$:
\begin{equation}
    \chi_{\text{unpol}}(\omega) = 45 \, \chi_1(\omega) + 7 \sum_{k=2}^{7} \chi_k(\omega),
\end{equation}
where $\chi_1(\omega)$ corresponds to the isotropic contribution, while $\chi_2(\omega), \dots, \chi_7(\omega)$ collectively account for the anisotropic invariant.

\bibliographystyle{apsrev4-2}

\bibliography{references}

@article{monacelli_hydrogen_2025,
	title = {Hydrogen bond symmetrization in high-pressure ice clathrates},
	volume = {112},
	doi = {10.1103/1cgl-mklx},
	number = {21},
	journal = {Physical Review B},
	author = {Monacelli, Lorenzo and {Maria Rescigno} and {Alasdair Nicholls} and {Umbertoluca Ranieri} and {Simone Di Cataldo} and {Livia Eleonora Bove}},
	year = {2025},
}

@article{fransson_dynasortool_2021,
	title = {dynasor—{A} {Tool} for {Extracting} {Dynamical} {Structure} {Factors} and {Current} {Correlation} {Functions} from {Molecular} {Dynamics} {Simulations}},
	volume = {4},
	copyright = {© 2021 The Authors. Advanced Theory and Simulations published by Wiley-VCH GmbH},
	issn = {2513-0390},
	url = {https://onlinelibrary.wiley.com/doi/abs/10.1002/adts.202000240},
	doi = {10.1002/adts.202000240},
	language = {en},
	number = {2},
	urldate = {2026-07-26},
	journal = {Advanced Theory and Simulations},
	author = {Fransson, Erik and Slabanja, Mattias and Erhart, Paul and Wahnström, Göran},
	year = {2021},
	pages = {2000240},
}

@article{eriksson_hiphive_2019-1,
	title = {The {Hiphive} {Package} for the {Extraction} of {High}-{Order} {Force} {Constants} by {Machine} {Learning}},
	volume = {2},
	copyright = {© 2019 WILEY-VCH Verlag GmbH \& Co. KGaA, Weinheim},
	issn = {2513-0390},
	url = {https://onlinelibrary.wiley.com/doi/abs/10.1002/adts.201800184},
	doi = {10.1002/adts.201800184},
	language = {en},
	number = {5},
	urldate = {2026-07-21},
	journal = {Advanced Theory and Simulations},
	author = {Eriksson, Fredrik and Fransson, Erik and Erhart, Paul},
	year = {2019},
	pages = {1800184},
}

@article{ranalli_temperature-dependent_2023,
	title = {Temperature-{Dependent} {Anharmonic} {Phonons} in {Quantum} {Paraelectric} {KTaO3} by {First} {Principles} and {Machine}-{Learned} {Force} {Fields}},
	volume = {6},
	copyright = {© 2023 The Authors. Advanced Quantum Technologies published by Wiley-VCH GmbH},
	issn = {2511-9044},
	doi = {10.1002/qute.202200131},
	number = {4},
	urldate = {2024-02-29},
	journal = {Advanced Quantum Technologies},
	author = {Ranalli, Luigi and Verdi, Carla and Monacelli, Lorenzo and Kresse, Georg and Calandra, Matteo and Franchini, Cesare},
	year = {2023},
	pages = {2200131},
}

@article{thomas_predicting_2010,
	title = {Predicting phonon dispersion relations and lifetimes from the spectral energy density},
	volume = {81},
	url = {https://link.aps.org/doi/10.1103/PhysRevB.81.081411},
	doi = {10.1103/PhysRevB.81.081411},
	number = {8},
	urldate = {2026-07-26},
	journal = {Physical Review B},
	publisher = {American Physical Society},
	author = {Thomas, John A. and Turney, Joseph E. and Iutzi, Ryan M. and Amon, Cristina H. and McGaughey, Alan J. H.},
	month = feb,
	year = {2010},
	pages = {081411},
}

@article{wu_lattice_2023,
	title = {Lattice {Thermal} {Conductivity} in {XMg2Sb2}({X} = {Ca} or {Mg}) {Compounds}: {Temperature} and {High}-{Order} {Anharmonicity} {Effect}},
	volume = {16},
	copyright = {http://creativecommons.org/licenses/by/3.0/},
	issn = {1996-1944},
	shorttitle = {Lattice {Thermal} {Conductivity} in {XMg2Sb2}({X} = {Ca} or {Mg}) {Compounds}},
	url = {https://www.mdpi.com/1996-1944/16/23/7349},
	doi = {10.3390/ma16237349},
	language = {en},
	number = {23},
	urldate = {2026-07-21},
	journal = {Materials},
	publisher = {Multidisciplinary Digital Publishing Institute},
	author = {Wu, Minghui and Yang, Hongping and Xie, Fengyan and Huang, Li},
	month = jan,
	year = {2023},
	pages = {7349},
}

@article{lin_first-principles_2026,
	title = {First-principles phonon physics using the {Pheasy} code},
	copyright = {2026 The Author(s)},
	issn = {2057-3960},
	url = {https://www.nature.com/articles/s41524-026-02163-1},
	doi = {10.1038/s41524-026-02163-1},
	language = {en},
	urldate = {2026-07-21},
	journal = {npj Computational Materials},
	publisher = {Nature Publishing Group},
	author = {Lin, Changpeng and Han, Jian and Xu, Ben and Marzari, Nicola},
	month = jun,
	year = {2026},
}

@article{togo_distributions_2015,
	title = {Distributions of phonon lifetimes in {Brillouin} zones},
	volume = {91},
	url = {https://link.aps.org/doi/10.1103/PhysRevB.91.094306},
	doi = {10.1103/PhysRevB.91.094306},
	number = {9},
	urldate = {2026-07-21},
	journal = {Physical Review B},
	publisher = {American Physical Society},
	author = {Togo, Atsushi and Chaput, Laurent and Tanaka, Isao},
	month = mar,
	year = {2015},
	pages = {094306},
}

@article{zhou_compressive_2019,
	title = {Compressive sensing lattice dynamics. {II}. {Efficient} phonon calculations and long-range interactions},
	volume = {100},
	url = {https://link.aps.org/doi/10.1103/PhysRevB.100.184309},
	doi = {10.1103/PhysRevB.100.184309},
	number = {18},
	urldate = {2026-07-21},
	journal = {Physical Review B},
	publisher = {American Physical Society},
	author = {Zhou, Fei and Sadigh, Babak and Åberg, Daniel and Xia, Yi and Ozoliņš, Vidvuds},
	month = nov,
	year = {2019},
	pages = {184309},
}

@article{zhou_lattice_2014,
	title = {Lattice {Anharmonicity} and {Thermal} {Conductivity} from {Compressive} {Sensing} of {First}-{Principles} {Calculations}},
	volume = {113},
	url = {https://link.aps.org/doi/10.1103/PhysRevLett.113.185501},
	doi = {10.1103/PhysRevLett.113.185501},
	number = {18},
	urldate = {2026-07-21},
	journal = {Physical Review Letters},
	publisher = {American Physical Society},
	author = {Zhou, Fei and Nielson, Weston and Xia, Yi and Ozoliņš, Vidvuds},
	month = oct,
	year = {2014},
	pages = {185501},
}

@misc{xia_quintic-anharmonicity-assisted_2026,
	title = {Quintic-{Anharmonicity}-{Assisted} {Three}-{Phonon} {Scattering}: {A} {Previously} {Overlooked} {Same}-{Order} {Channel} to {Four}-{Phonon} {Scattering}},
	shorttitle = {Quintic-{Anharmonicity}-{Assisted} {Three}-{Phonon} {Scattering}},
	url = {http://arxiv.org/abs/2606.24409},
	doi = {10.48550/arXiv.2606.24409},
	urldate = {2026-07-20},
	publisher = {arXiv},
	author = {Xia, Yi},
	month = jun,
	year = {2026},
	note = {arXiv:2606.24409 [cond-mat.mtrl-sci]},
}

@article{siciliano_wigner_2023,
	title = {Wigner {Gaussian} dynamics: {Simulating} the anharmonic and quantum ionic motion},
	volume = {107},
	shorttitle = {Wigner {Gaussian} dynamics},
	url = {https://link.aps.org/doi/10.1103/PhysRevB.107.174307},
	doi = {10.1103/PhysRevB.107.174307},
	number = {17},
	urldate = {2023-09-07},
	journal = {Physical Review B},
	publisher = {American Physical Society},
	author = {Siciliano, Antonio and Monacelli, Lorenzo and Caldarelli, Giovanni and Mauri, Francesco},
	month = may,
	year = {2023},
	pages = {174307},
}

@article{lihm_gaussian_2021,
	title = {Gaussian time-dependent variational principle for the finite-temperature anharmonic lattice dynamics},
	volume = {3},
	url = {https://link.aps.org/doi/10.1103/PhysRevResearch.3.L032017},
	doi = {10.1103/PhysRevResearch.3.L032017},
	number = {3},
	urldate = {2026-03-21},
	journal = {Physical Review Research},
	publisher = {American Physical Society},
	author = {Lihm, Jae-Mo and Park, Cheol-Hwan},
	month = jul,
	year = {2021},
	pages = {L032017},
}

@article{libbi_nonequilibrium_2025,
	title = {Nonequilibrium quantum dynamics in {SrTiO3} under impulsive {THz} radiation with machine learning},
	volume = {11},
	url = {https://www.science.org/doi/10.1126/sciadv.adw1634},
	doi = {10.1126/sciadv.adw1634},
	number = {37},
	urldate = {2026-03-21},
	journal = {Science Advances},
	publisher = {American Association for the Advancement of Science},
	author = {Libbi, Francesco and Johansson, Anders and Kozinsky, Boris and Monacelli, Lorenzo},
	month = sep,
	year = {2025},
	pages = {eadw1634},
}

@article{monacelli_analyzing_2025,
	title = {Analyzing the anharmonic phonon spectrum: {Self}-consistent approximation and temperature-dependent effective potential methods},
	volume = {112},
	shorttitle = {Analyzing the anharmonic phonon spectrum},
	url = {https://link.aps.org/doi/10.1103/8611-5k5v},
	doi = {10.1103/8611-5k5v},
	number = {1},
	urldate = {2025-07-25},
	journal = {Physical Review B},
	publisher = {American Physical Society},
	author = {Monacelli, Lorenzo},
	month = jul,
	year = {2025},
	pages = {014109},
}

@article{knoop_tdep_2024,
	title = {{TDEP}: {Temperature} {Dependent} {Effective} {Potentials}},
	volume = {9},
	issn = {2475-9066},
	shorttitle = {{TDEP}},
	url = {https://joss.theoj.org/papers/10.21105/joss.06150},
	doi = {10.21105/joss.06150},
	language = {en},
	number = {94},
	urldate = {2025-04-17},
	journal = {Journal of Open Source Software},
	author = {Knoop, Florian and Shulumba, Nina and Castellano, Aloïs and Batista, J. P. Alvarinhas and Farris, Roberta and Verstraete, Matthieu J. and Heine, Matthew and Broido, David and Kim, Dennis S. and Klarbring, Johan and Abrikosov, Igor A. and Simak, Sergei I. and Hellman, Olle},
	month = feb,
	year = {2024},
	pages = {6150},
}

@article{ribeiro_strong_2018,
	title = {Strong anharmonicity in the phonon spectra of {PbTe} and {SnTe} from first principles},
	volume = {97},
	url = {https://doi.org/10.1103%2Fphysrevb.97.014306},
	doi = {10.1103/physrevb.97.014306},
	number = {1},
	journal = {Physical Review B},
	publisher = {American Physical Society (APS)},
	author = {Ribeiro, Guilherme A. S. and Paulatto, Lorenzo and Bianco, Raffaello and Errea, Ion and Mauri, Francesco and Calandra, Matteo},
	month = jan,
	year = {2018},
	pages = {014306},
}

@article{romero_thermal_2015,
	title = {Thermal conductivity in {PbTe} from first principles},
	volume = {91},
	url = {https://link.aps.org/doi/10.1103/PhysRevB.91.214310},
	doi = {10.1103/PhysRevB.91.214310},
	number = {21},
	urldate = {2025-02-20},
	journal = {Physical Review B},
	publisher = {American Physical Society},
	author = {Romero, A. H. and Gross, E. K. U. and Verstraete, M. J. and Hellman, Olle},
	month = jun,
	year = {2015},
	pages = {214310},
}

@article{cherubini_quantum_2024,
	title = {Quantum effects in {H}-bond symmetrization and in thermodynamic properties of high pressure ice},
	volume = {110},
	url = {https://link.aps.org/doi/10.1103/PhysRevB.110.014112},
	doi = {10.1103/PhysRevB.110.014112},
	number = {1},
	urldate = {2024-11-20},
	journal = {Physical Review B},
	publisher = {American Physical Society},
	author = {Cherubini, Marco and Monacelli, Lorenzo and Yang, Bingjia and Car, Roberto and Casula, Michele and Mauri, Francesco},
	month = jul,
	year = {2024},
	pages = {014112},
}

@article{shin_quantum_2021,
	title = {Quantum paraelectric phase of {SrTiO3} from first principles},
	volume = {104},
	url = {https://link.aps.org/doi/10.1103/PhysRevB.104.L060103},
	doi = {10.1103/PhysRevB.104.L060103},
	number = {6},
	urldate = {2024-02-23},
	journal = {Physical Review B},
	publisher = {American Physical Society},
	author = {Shin, Dongbin and Latini, Simone and Schäfer, Christian and Sato, Shunsuke A. and De Giovannini, Umberto and Hübener, Hannes and Rubio, Angel},
	month = aug,
	year = {2021},
	pages = {L060103},
}

@article{xian_li_terahertz_2019,
	title = {Terahertz field–induced ferroelectricity in quantum paraelectric {SrTiO3}},
	volume = {364},
	url = {https://www.science.org/doi/10.1126/science.aaw4913},
	doi = {10.1126/science.aaw4913},
	urldate = {2024-02-27},
	journal = {Science},
	author = {Xian Li and Tian Qiu and Jiahao Zhang and Edoardo Baldini and Jian Lu and Andrew M Rappe and Keith A Nelson},
	year = {2019},
	pages = {1079},
}

@article{SkyZhou2020,
	title = {Theory of the thickness dependence of the charge density wave transition in 1 {T}-{TiTe2}},
	volume = {7},
	copyright = {All rights reserved},
	url = {https://doi.org/10.10882F2053-15832Fabae7a},
	doi = {10.1088/2053-1583/abae7a},
	number = {4},
	journal = {2D Materials},
	publisher = {IOP Publishing},
	author = {Zhou, Jianqiang Sky and Bianco, Raffaello and Monacelli, Lorenzo and Errea, Ion and Mauri, Francesco and Calandra, Matteo},
	month = sep,
	year = {2020},
	pages = {045032},
}

@article{bianco_quantum_2019,
	title = {Quantum enhancement of charge density wave in {NbS2} in the two-dimensional limit},
	volume = {19},
	copyright = {All rights reserved},
	url = {https://doi.org/10.10212Facs.nanolett.9b00504},
	doi = {10.1021/acs.nanolett.9b00504},
	number = {5},
	journal = {Nano Letters},
	publisher = {American Chemical Society (ACS)},
	author = {Bianco, Raffaello and Errea, Ion and Monacelli, Lorenzo and Calandra, Matteo and Mauri, Francesco},
	month = apr,
	year = {2019},
	pages = {3098--3103},
}

@article{cheng_terahertz-driven_2023,
	title = {Terahertz-{Driven} {Local} {Dipolar} {Correlation} in a {Quantum} {Paraelectric}},
	volume = {130},
	issn = {0031-9007, 1079-7114},
	url = {https://link.aps.org/doi/10.1103/PhysRevLett.130.126902},
	doi = {10.1103/PhysRevLett.130.126902},
	number = {12},
	urldate = {2024-02-23},
	journal = {Physical Review Letters},
	author = {Cheng, Bing and Kramer, Patrick L. and Shen, Zhi-Xun and Hoffmann, Matthias C.},
	month = mar,
	year = {2023},
	pages = {126902},
}

@article{monacelli_quantum_2023,
	title = {Quantum phase diagram of high-pressure hydrogen},
	volume = {19},
	copyright = {2023 The Author(s), under exclusive licence to Springer Nature Limited},
	issn = {1745-2481},
	url = {https://www.nature.com/articles/s41567-023-01960-5},
	doi = {10.1038/s41567-023-01960-5},
	number = {6},
	urldate = {2024-08-28},
	journal = {Nature Physics},
	publisher = {Nature Publishing Group},
	author = {Monacelli, Lorenzo and Casula, Michele and Nakano, Kousuke and Sorella, Sandro and Mauri, Francesco},
	month = jun,
	year = {2023},
	pages = {845--850},
}

@article{aseginolaza_bending_2024,
	title = {Bending rigidity, sound propagation and ripples in flat graphene},
	volume = {20},
	copyright = {2024 The Author(s), under exclusive licence to Springer Nature Limited},
	issn = {1745-2481},
	url = {https://www.nature.com/articles/s41567-024-02441-z},
	doi = {10.1038/s41567-024-02441-z},
	language = {en},
	number = {8},
	urldate = {2024-08-28},
	journal = {Nature Physics},
	publisher = {Nature Publishing Group},
	author = {Aseginolaza, Unai and Diego, Josu and Cea, Tommaso and Bianco, Raffaello and Monacelli, Lorenzo and Libbi, Francesco and Calandra, Matteo and Bergara, Aitor and Mauri, Francesco and Errea, Ion},
	month = aug,
	year = {2024},
	pages = {1288--1293},
}

@misc{ranalli_electron_2024,
	title = {Electron {Mobilities} in {SrTiO}\$\_3\$ and {KTaO}\$\_3\$: {Role} of {Phonon} {Anharmonicity}, {Mass} {Renormalization} and {Disorder}},
	shorttitle = {Electron {Mobilities} in {SrTiO}\$\_3\$ and {KTaO}\$\_3\$},
	url = {http://arxiv.org/abs/2407.18771},
	doi = {10.48550/arXiv.2407.18771},
	urldate = {2024-08-28},
	publisher = {arXiv},
	author = {Ranalli, Luigi and Verdi, Carla and Zacharias, Marios and Evens, Jacky and Giustino, Feliciano and Franchini, Cesare},
	month = jul,
	year = {2024},
	note = {arXiv:2407.18771 [cond-mat]},
}

@article{souvatzis_entropy_2008,
	title = {Entropy {Driven} {Stabilization} of {Energetically} {Unstable} {Crystal} {Structures} {Explained} from {First} {Principles} {Theory}},
	volume = {100},
	url = {https://link.aps.org/doi/10.1103/PhysRevLett.100.095901},
	doi = {10.1103/PhysRevLett.100.095901},
	number = {9},
	urldate = {2024-06-18},
	journal = {Physical Review Letters},
	publisher = {American Physical Society},
	author = {Souvatzis, P. and Eriksson, O. and Katsnelson, M. I. and Rudin, S. P.},
	month = mar,
	year = {2008},
	pages = {095901},
}

@article{dangic_molecular_2022,
	title = {Molecular dynamics simulation of the ferroelectric phase transition in {GeTe}: {Displacive} or order-disorder character},
	volume = {106},
	shorttitle = {Molecular dynamics simulation of the ferroelectric phase transition in {GeTe}},
	url = {https://link.aps.org/doi/10.1103/PhysRevB.106.134113},
	doi = {10.1103/PhysRevB.106.134113},
	number = {13},
	urldate = {2024-06-15},
	journal = {Physical Review B},
	publisher = {American Physical Society},
	author = {Dangić, {{\fontencoding{T1}\selectfont\DJ}}or{{\fontencoding{T1}\selectfont\dj}}e and Fahy, Stephen and Savić, Ivana},
	month = oct,
	year = {2022},
	pages = {134113},
}

@article{bottin_-tdep_2020,
	title = {a-{TDEP}: {Temperature} {Dependent} {Effective} {Potential} for {Abinit} – {Lattice} dynamic properties including anharmonicity},
	volume = {254},
	issn = {0010-4655},
	shorttitle = {a-{TDEP}},
	url = {https://www.sciencedirect.com/science/article/pii/S0010465520301156},
	doi = {10.1016/j.cpc.2020.107301},
	urldate = {2024-04-19},
	journal = {Computer Physics Communications},
	author = {Bottin, François and Bieder, Jordan and Bouchet, Johann},
	month = sep,
	year = {2020},
	pages = {107301},
}

@article{castellano_mode-coupling_2023,
	title = {Mode-coupling theory of lattice dynamics for classical and quantum crystals},
	volume = {159},
	issn = {0021-9606},
	url = {https://doi.org/10.1063/5.0174255},
	doi = {10.1063/5.0174255},
	number = {23},
	urldate = {2024-04-17},
	journal = {The Journal of Chemical Physics},
	author = {Castellano, Aloïs and Batista, J. P. Alvarinhas and Verstraete, Matthieu J.},
	month = dec,
	year = {2023},
	pages = {234501},
}

@article{monacelli_first-principles_2023,
	title = {First-{Principles} {Thermodynamics} of {CsSnI3}},
	volume = {35},
	copyright = {All rights reserved},
	issn = {0897-4756},
	url = {https://doi.org/10.1021/acs.chemmater.2c03475},
	doi = {10.1021/acs.chemmater.2c03475},
	number = {4},
	urldate = {2024-04-11},
	journal = {Chemistry of Materials},
	publisher = {American Chemical Society},
	author = {Monacelli, Lorenzo and Marzari, Nicola},
	month = feb,
	year = {2023},
	pages = {1702--1709},
}

@article{nova_metastable_2019,
	title = {Metastable ferroelectricity in optically strained {SrTiO3}},
	volume = {364},
	url = {https://www.science.org/doi/10.1126/science.aaw4911},
	doi = {10.1126/science.aaw4911},
	number = {6445},
	urldate = {2024-02-21},
	journal = {Science},
	publisher = {American Association for the Advancement of Science},
	author = {Nova, T. F. and Disa, A. S. and Fechner, M. and Cavalleri, A.},
	month = jun,
	year = {2019},
	pages = {1075--1079},
}

@article{verdi_quantum_2023,
	title = {Quantum paraelectricity and structural phase transitions in strontium titanate beyond density functional theory},
	volume = {7},
	url = {https://link.aps.org/doi/10.1103/PhysRevMaterials.7.L030801},
	doi = {10.1103/PhysRevMaterials.7.L030801},
	number = {3},
	urldate = {2024-02-19},
	journal = {Physical Review Materials},
	publisher = {American Physical Society},
	author = {Verdi, Carla and Ranalli, Luigi and Franchini, Cesare and Kresse, Georg},
	month = mar,
	year = {2023},
	pages = {L030801},
}

@article{cochran_dielectric_1962,
	title = {Dielectric constants and lattice vibrations},
	volume = {23},
	issn = {0022-3697},
	url = {https://www.sciencedirect.com/science/article/pii/0022369762900847},
	doi = {10.1016/0022-3697(62)90084-7},
	number = {5},
	urldate = {2024-02-08},
	journal = {Journal of Physics and Chemistry of Solids},
	author = {Cochran, W. and Cowley, R. A.},
	month = may,
	year = {1962},
	pages = {447--450},
}

@article{cochran_crystal_1960,
	title = {Crystal stability and the theory of ferroelectricity},
	copyright = {Copyright Taylor and Francis Group, LLC},
	url = {https://www.tandfonline.com/doi/abs/10.1080/00018736000101229},
	doi = {10.1080/00018736000101229},
	language = {EN},
	urldate = {2024-02-08},
	journal = {Advances in Physics},
	publisher = {Taylor \& Francis Group},
	author = {Cochran, W.},
	month = oct,
	year = {1960},
}

@article{dangic_origin_2021,
	title = {The origin of the lattice thermal conductivity enhancement at the ferroelectric phase transition in {GeTe}},
	volume = {7},
	copyright = {2021 The Author(s)},
	issn = {2057-3960},
	url = {https://www.nature.com/articles/s41524-021-00523-7},
	doi = {10.1038/s41524-021-00523-7},
	language = {en},
	number = {1},
	urldate = {2023-10-13},
	journal = {npj Computational Materials},
	publisher = {Nature Publishing Group},
	author = {Dangić, {{\fontencoding{T1}\selectfont\DJ}}or{{\fontencoding{T1}\selectfont\dj}}e and Hellman, Olle and Fahy, Stephen and Savić, Ivana},
	month = apr,
	year = {2021},
	note = {Number: 1},
	pages = {1--8},
}

@misc{melnikov_anharmonic_2023,
	title = {Anharmonic coherent dynamics of the soft phonon mode of a {PbTe} crystal},
	url = {http://arxiv.org/abs/2309.06549},
	urldate = {2023-09-22},
	publisher = {arXiv},
	author = {Melnikov, A. A. and Selivanov, Yu G. and Chekalin, S. V.},
	month = sep,
	year = {2023},
	note = {arXiv:2309.06549 [cond-mat, physics:physics]},
}

@article{monacelli_time-dependent_2021,
	title = {Time-dependent self-consistent harmonic approximation: {Anharmonic} nuclear quantum dynamics and time correlation functions},
	volume = {103},
	copyright = {All rights reserved},
	url = {https://doi.org/10.1103/physrevb.103.104305},
	doi = {10.1103/physrevb.103.104305},
	number = {10},
	journal = {Physical Review B},
	publisher = {American Physical Society (APS)},
	author = {Monacelli, Lorenzo and Mauri, Francesco},
	month = mar,
	year = {2021},
	pages = {104305},
}

@article{monacelli_stochastic_2021,
	title = {The stochastic self-consistent harmonic approximation: calculating vibrational properties of materials with full quantum and anharmonic effects},
	volume = {33},
	copyright = {All rights reserved},
	url = {https://doi.org/10.1088/1361-648x/ac066b},
	doi = {10.1088/1361-648x/ac066b},
	number = {36},
	journal = {Journal of Physics: Condensed Matter},
	publisher = {IOP Publishing},
	author = {Monacelli, Lorenzo and Bianco, Raffaello and Cherubini, Marco and Calandra, Matteo and Errea, Ion and Mauri, Francesco},
	month = jul,
	year = {2021},
	pages = {363001},
}

@article{aseginolaza_strong_2019,
	title = {Strong anharmonicity and high thermoelectric efficiency in high-temperature {SnS} from first principles},
	volume = {100},
	url = {https://doi.org/10.1103%2Fphysrevb.100.214307},
	doi = {10.1103/physrevb.100.214307},
	number = {21},
	journal = {Physical Review B},
	publisher = {American Physical Society (APS)},
	author = {Aseginolaza, Unai and Bianco, Raffaello and Monacelli, Lorenzo and Paulatto, Lorenzo and Calandra, Matteo and Mauri, Francesco and Bergara, Aitor and Errea, Ion},
	month = dec,
	year = {2019},
	pages = {214307},
}

@article{bianco_second-order_2017,
	title = {Second-order structural phase transitions, free energy curvature, and temperature-dependent anharmonic phonons in the self-consistent harmonic approximation: {Theory} and stochastic implementation},
	volume = {96},
	url = {https://doi.org/10.1103%2Fphysrevb.96.014111},
	doi = {10.1103/physrevb.96.014111},
	number = {1},
	journal = {Physical Review B},
	publisher = {American Physical Society (APS)},
	author = {Bianco, Raffaello and Errea, Ion and Paulatto, Lorenzo and Calandra, Matteo and Mauri, Francesco},
	month = jul,
	year = {2017},
	pages = {014111},
}

@article{tadano_self-consistent_2015,
	title = {Self-consistent phonon calculations of lattice dynamical properties in cubic {SrTiO3} with first-principles anharmonic force constants},
	volume = {92},
	number = {5},
	journal = {Physical Review B},
	publisher = {American Physical Society},
	author = {Tadano, Terumasa and Tsuneyuki, Shinji},
	month = aug,
	year = {2015},
	note = {Number of pages: 10},
	pages = {054301},
}

@article{errea_quantum_2016,
	title = {Quantum hydrogen-bond symmetrization in the superconducting hydrogen sulfide system},
	volume = {532},
	url = {https://doi.org/10.1038%2Fnature17175},
	doi = {10.1038/nature17175},
	number = {7597},
	journal = {Nature},
	publisher = {Springer Science and Business Media LLC},
	author = {Errea, Ion and Calandra, Matteo and Pickard, Chris J. and Nelson, Joseph R. and Needs, Richard J. and Li, Yinwei and Liu, Hanyu and Zhang, Yunwei and Ma, Yanming and Mauri, Francesco},
	month = mar,
	year = {2016},
	pages = {81--84},
}

@article{errea_quantum_2020,
	title = {Quantum crystal structure in the 250-kelvin superconducting lanthanum hydride},
	volume = {578},
	url = {https://doi.org/10.1038%2Fs41586-020-1955-z},
	doi = {10.1038/s41586-020-1955-z},
	number = {7793},
	journal = {Nature},
	publisher = {Springer Science and Business Media LLC},
	author = {Errea, Ion and Belli, Francesco and Monacelli, Lorenzo and Sanna, Antonio and Koretsune, Takashi and Tadano, Terumasa and Bianco, Raffaello and Calandra, Matteo and Arita, Ryotaro and Mauri, Francesco and Flores-Livas, José A.},
	month = feb,
	year = {2020},
	pages = {66--69},
}

@article{monacelli_pressure_2018,
	title = {Pressure and stress tensor of complex anharmonic crystals within the stochastic self-consistent harmonic approximation},
	volume = {98},
	copyright = {All rights reserved},
	url = {https://doi.org/10.1103/physrevb.98.024106},
	doi = {10.1103/physrevb.98.024106},
	number = {2},
	journal = {Physical Review B},
	publisher = {American Physical Society (APS)},
	author = {Monacelli, Lorenzo and Errea, Ion and Calandra, Matteo and Mauri, Francesco},
	month = jul,
	year = {2018},
	pages = {024106},
}

@article{aseginolaza_phonon_2019,
	title = {Phonon collapse and second-order phase transition in thermoelectric {SnSe}},
	volume = {122},
	url = {https://doi.org/10.1103/physrevlett.122.075901},
	doi = {10.1103/physrevlett.122.075901},
	number = {7},
	journal = {Physical Review Letters},
	publisher = {American Physical Society (APS)},
	author = {Aseginolaza, Unai and Bianco, Raffaello and Monacelli, Lorenzo and Paulatto, Lorenzo and Calandra, Matteo and Mauri, Francesco and Bergara, Aitor and Errea, Ion},
	month = feb,
	year = {2019},
	pages = {075901},
}

@article{hellman_lattice_2011,
	title = {Lattice dynamics of anharmonic solids from first principles},
	volume = {84},
	url = {https://doi.org/10.11032Fphysrevb.84.180301},
	doi = {10.1103/physrevb.84.180301},
	number = {18},
	journal = {Physical Review B},
	publisher = {American Physical Society (APS)},
	author = {Hellman, O. and Abrikosov, I. A. and Simak, S. I.},
	month = nov,
	year = {2011},
	pages = {180301},
}

@article{monacelli_black_2020,
	title = {Black metal hydrogen above 360 {GPa} driven by proton quantum fluctuations},
	volume = {17},
	url = {https://doi.org/10.1038%2Fs41567-020-1009-3},
	doi = {10.1038/s41567-020-1009-3},
	number = {1},
	journal = {Nature Physics},
	publisher = {Springer Science and Business Media LLC},
	author = {Monacelli, Lorenzo and Errea, Ion and Calandra, Matteo and Mauri, Francesco},
	month = sep,
	year = {2020},
	pages = {63--67},
}

@article{errea_anharmonic_2014,
	title = {Anharmonic free energies and phonon dispersions from the stochastic self-consistent harmonic approximation: {Application} to platinum and palladium hydrides},
	volume = {89},
	url = {https://doi.org/10.1103/physrevb.89.064302},
	doi = {10.1103/physrevb.89.064302},
	number = {6},
	journal = {Physical Review B},
	publisher = {American Physical Society (APS)},
	author = {Errea, Ion and Calandra, Matteo and Mauri, Francesco},
	month = feb,
	year = {2014},
	pages = {064302},
}

@article{tadano_anharmonic_2014,
	title = {Anharmonic force constants extracted from first-principles molecular dynamics: applications to heat transfer simulations},
	volume = {26},
	url = {https://doi.org/10.1088%2F0953-8984%2F26%2F22%2F225402},
	doi = {10.1088/0953-8984/26/22/225402},
	number = {22},
	journal = {Journal of Physics: Condensed Matter},
	publisher = {IOP Publishing},
	author = {Tadano, T and Gohda, Y and Tsuneyuki, S},
	month = may,
	year = {2014},
	pages = {225402},
}

@article{phono3py,
	title = {Distributions of phonon lifetimes in {Brillouin} zones},
	volume = {91},
	doi = {10.1103/PhysRevB.91.094306},
	number = {9},
	journal = {Physical Review B},
	publisher = {American Physical Society},
	author = {Togo, Atsushi and Chaput, Laurent and Tanaka, Isao},
	month = mar,
	year = {2015},
	note = {Number of pages: 31},
	pages = {094306},
}

@article{Borinaga2017,
	title = {Anharmonic effects in atomic hydrogen: {Superconductivity} and lattice dynamical stability},
	volume = {93},
	url = {https://link.aps.org/doi/10.1103/PhysRevB.93.174308},
	doi = {10.1103/PhysRevB.93.174308},
	number = {17},
	journal = {Physical Review B},
	publisher = {American Physical Society},
	author = {Borinaga, Miguel and Errea, Ion and Calandra, Matteo and Mauri, Francesco and Bergara, Aitor},
	month = may,
	year = {2016},
	note = {Number of pages: 8},
	pages = {174308},
}

@article{QSCALID,
	title = {Quantum self-consistent ab-initio lattice dynamics},
	volume = {263},
	url = {https://doi.org/10.1016%2Fj.cpc.2021.107945},
	doi = {10.1016/j.cpc.2021.107945},
	journal = {Computer Physics Communications},
	publisher = {Elsevier BV},
	author = {van Roekeghem, Ambroise and Carrete, Jesús and Mingo, Natalio},
	month = jun,
	year = {2021},
	pages = {107945},
}

@article{Kaiser2021,
	title = {First-principles molecular dynamics in metal-halide perovskites: {Contrasting} generalized gradient approximation and hybrid functionals},
	volume = {12},
	url = {https://doi.org/10.1021/acs.jpclett.1c03428},
	doi = {10.1021/acs.jpclett.1c03428},
	number = {49},
	journal = {The Journal of Physical Chemistry Letters},
	publisher = {American Chemical Society (ACS)},
	author = {Kaiser, Waldemar and Carignano, Marcelo and Alothman, Asma A. and Mosconi, Edoardo and Kachmar, Ali and Goddard, William A. and Angelis, Filippo De},
	month = dec,
	year = {2021},
	pages = {11886--11893},
}

@article{Diego2021,
	title = {van der {Waals} driven anharmonic melting of the {3D} charge density wave in {VSe2}},
	volume = {12},
	copyright = {All rights reserved},
	url = {https://doi.org/10.10382Fs41467-020-20829-2},
	doi = {10.1038/s41467-020-20829-2},
	number = {1},
	journal = {Nature Communications},
	publisher = {Springer Science and Business Media LLC},
	author = {Diego, Josu and Said, A. H. and Mahatha, S. K. and Bianco, Raffaello and Monacelli, Lorenzo and Calandra, Matteo and Mauri, Francesco and Rossnagel, K. and Errea, Ion and Blanco-Canosa, S.},
	month = jan,
	year = {2021},
	pages = {598},
}

@article{SCALID,
	title = {The self-consistent ab initio lattice dynamical method},
	volume = {44},
	url = {https://doi.org/10.1016%2Fj.commatsci.2008.06.016},
	doi = {10.1016/j.commatsci.2008.06.016},
	number = {3},
	journal = {Computational Materials Science},
	publisher = {Elsevier BV},
	author = {Souvatzis, P. and Eriksson, O. and Katsnelson, M.I. and Rudin, S.P.},
	month = jan,
	year = {2009},
	pages = {888--894},
}

@article{Chung2012,
	title = {{CsSnI3}: {Semiconductor} or metal? {High} electrical conductivity and strong near-infrared photoluminescence from a single material. {High} hole mobility and phase-transitions},
	volume = {134},
	url = {https://doi.org/10.1021/ja301539s},
	doi = {10.1021/ja301539s},
	number = {20},
	journal = {Journal of the American Chemical Society},
	publisher = {American Chemical Society (ACS)},
	author = {Chung, In and Song, Jung-Hwan and Im, Jino and Androulakis, John and Malliakas, Christos D. and Li, Hao and Freeman, Arthur J. and Kenney, John T. and Kanatzidis, Mercouri G.},
	month = may,
	year = {2012},
	pages = {8579--8587},
}

\end{document}